\UseRawInputEncoding
\documentclass[5p]{elsarticle}

\usepackage{tcolorbox}
\usepackage[T1]{fontenc}
\usepackage[utf8]{inputenc}
\usepackage{amsmath}  
\usepackage{amssymb}
\usepackage{graphicx}
\usepackage{caption}
\usepackage{subcaption} 
\usepackage{amsmath}
\usepackage{mathtools}
\usepackage[version=4]{mhchem}
\usepackage{enumitem}
\usepackage{upgreek}
\usepackage{array}
\usepackage{soul}
\usepackage{tabularx}
\usepackage{threeparttable}
\usepackage[
list-final-separator={, \text{and} },
retain-explicit-plus,
]{siunitx}
\DeclareSIUnit\angstrom{\text{Å}}
\usepackage{tablefootnote}
\usepackage{multirow}
\usepackage{xcolor, color}
\sethlcolor{yellow}
\usepackage{booktabs, caption, makecell}
\usepackage[colorlinks,citecolor=red,urlcolor=blue,bookmarks=false,hypertexnames=true]{hyperref} 
\hypersetup{colorlinks = true,linkcolor = blue,anchorcolor =red,citecolor = blue,filecolor = red,urlcolor = red,
           pdfauthor=author}
\newcolumntype{C}[1]{>{\centering\let\newline\\\arraybackslash\hspace{0pt}}m{#1}}
\DeclareUnicodeCharacter{2212}{\textendash}
\DeclareUnicodeCharacter{03B1}{$\alpha$}
\DeclareUnicodeCharacter{03B2}{$\beta$}
\DeclareUnicodeCharacter{03B4}{$\delta$}
\DeclareUnicodeCharacter{00E1}{\'{a}}
\DeclareUnicodeCharacter{223C}{$\sim$}
\biboptions{sort&compress}
\definecolor{lightgreenyellow}{RGB}{255, 255, 200}

\begin{document}
\sloppy  
\begin{frontmatter}

\title{Machine Learning Potentials for Hydrogen Absorption in \texorpdfstring{TiCr\textsubscript{2}}{TiCr2} Laves Phases}

\affiliation[inst1]{
organization={Institute for Materials Science, University of Stuttgart},
addressline={Pfaffenwaldring 55},
postcode={70569},
city={Stuttgart},
country={Germany},
}
\affiliation[inst2]{
organization={Interdisciplinary Centre for Advanced Materials Simulation (ICAMS), Ruhr-Universität Bochum},
postcode={44801},
city={Bochum},
country={Germany},
}
           
\author[inst1]{Pranav Kumar\corref{cor1}}
\ead{pranav.kumar@imw.uni-stuttgart.de}
\author[inst1,inst2]{Fritz Körmann}
\author[inst1]{Blazej Grabowski}
\author[inst1]{Yuji Ikeda\corref{cor1}}
\ead{yuji.ikeda@imw.uni-stuttgart.de}
\cortext[cor1]
{Corresponding authors}            
\noindent

\begin{abstract}
The energetics of hydrogen absorption in C15 cubic and C14 hexagonal \ce{TiCr2H_x} Laves phases is investigated for $0 < x \le 6$ with density functional theory (DFT) and machine learning interatomic potentials (MLIPs).
The MLIPs are trained with configurations generated through a series of active-learning schemes.
Basin-hopping Monte Carlo (BHMC) simulations based on the MLIPs predict minimum-energy hydrogen configurations, along with enthalpies of formation and hydrogen orderings.
The obtained phase transformations at \qty{0}{K} agree well with the experiments at low temperatures.
The hydrogen solubility limits in the low-concentration $\alpha$ phases at \qty{0}{K} are predicted to be $x = 1.0$ and $x = 1.5$ for the C15 and the C14 phases, respectively.
At these concentrations, C15 \ce{TiCr2H} shows the $Cc$ monoclinic symmetry, while C14 \ce{TiCr2H_{1.5}} shows the $Ama2$ orthorhombic symmetry, both of which have not been reported for this system.
The first and the second hydride phases, i.e., $\beta$ and $\beta'$, at 0\,K are found around $x = 3$ and $x = 4$, respectively, for both the C15 and the C14 phases.
In the second-hydride $\beta'$ phases, C15 \ce{TiCr2H4} shows the $I4_1/a$ tetragonal symmetry, while C14 \ce{TiCr2H4} shows the $R\bar3c$ rhombohedral symmetry.
Hydrogen repulsion are found to extend to edge-sharing interstices, affecting the hydrogen ordering.
Furthermore, the $6h_2$ \ce{A2B2} interstices are found to be energetically substantially more preferable for C14 \ce{TiCr2H_x} than the other \ce{A2B2} interstices at low hydrogen concentrations, influencing the hydrogen-occupation trend.
\par
    \begin{tcolorbox}[
    colback=lightgreenyellow,
    colframe=white,
    boxrule=0.5pt,
    sharp corners,
    width=\textwidth,  
    left=2mm, right=2mm, top=1mm, bottom=1mm
]
        Published in: Acta Materialia 297 (2025) 121319, 
        \url{https://doi.org/10.1016/j.actamat.2025.121319}
\end{tcolorbox}
\end{abstract}







\end{frontmatter}




\section{Introduction}
\label{sec:Introduction}

Hydrogen is considered a cleaner and greener alternative to conventional energy sources~\cite{acar_review_2019}. Various hydrogen-storage methods exist, including physical approaches such as high-pressure gaseous storage and liquid storage, as well as solid-state storage techniques that utilize, e.g., metal hydrides~\cite{shahi_perspectives_2023,marques_review_2021} and nanoporous materials~\cite{murray_hydrogen_2009,chen_computational_2024,bobbitt_molecular_2019}.
Solid-state storage has safety advantages and is thus a promising alternative to the more established methods based on gaseous hydrogen~\cite{walker_solid-state_2008}.

The \ce{AB2} Laves-phase alloys are particularly interesting hydrogen-storage materials~\cite{sinha_hydrogen_1985,lin_recent_2022,yartys_laves_2022,chanchetti_scientometric_2020, kandavel_improvement_2008, ponsoni_design_2022}, offering good hydrogen capacity as well as other suitable functional and structural properties \cite{stein_laves_2021}. These materials exhibit wide chemical diversity, with the A site typically occupied by elements having strong hydrogen binding tendencies such as \ce{Mg}, \ce{Ca}, \ce{Ti}, and \ce{Zr} and with the B site occupied by 3d transition metals such as \ce{V}, \ce{Cr}, \ce{Mn}, \ce{Fe}, and \ce{Ni}.
Further, the possibility of generating an extensive range of mixtures both on the A and the B sites strongly increases the number of hydrogen-storage candidates, with the TiZrCrMnFeNi high-entropy alloy being a prominent example~\cite{andrade_crystal_2023,edalati_reversible_2020,dangwal_machine_2024,floriano_hydrogen_2020}.

Electronic-structure studies based on density functional theory (DFT) have provided valuable insights into the hydrogen absorption and diffusion in the \ce{AB2} Laves phases~\cite{li_hydrogen_2009,
gesari_hydrogen_2010,
burr_hydrogen_2013,
radakovic_hydrogen_2013,
radakovic_interstitial_2014,
merlino_dft_2016,
robina_hydrogen_2018,
song_first-principles_2021,
jiang_influence_2022,
mohammadi_high-entropy_2022,
ding_interface_2023,
loh_substitutional_2023,
strozi_tuning_2023,
somo_thermodynamic_2024}.
Most of the studies have focused on the binding energy and migration barriers of hydrogen in the dilute limit.
Hydrogen absorption or diffusion at finite hydrogen concentrations was occasionally considered, but only under very restricted conditions.
For example, Li~\textit{et~al.}~\cite{li_hydrogen_2009} investigated the hydrogen binding energy for C15 \ce{TiCr2H_x} based on DFT across a wide range of hydrogen concentrations ($0.5 \le x \le 12$).
They considered a geometrical criterion to maximize the distances between hydrogen atoms.
However, their analysis was limited to a single configuration at each concentration, which is not sufficient to provide conclusive insights into the distribution of hydrogen.

The limitations in the previous studies employing DFT is attributed to its immense computational cost.
Further progress requires alternative strategies to rigorously yet efficiently study finite hydrogen concentrations. An alternative is offered by interatomic potentials, which broadly fall into two classes. The class of physics-based semi-empirical interatomic potentials features good transferability. Various such potentials were proposed for the study of metal--hydrogen interactions \cite{lee_comparative_2014,mason_empirical_2023,tehranchi_atomistic_2017,smirnova_new_2018,bonny_binding_2014,lee_modified_2007,meng_general-purpose_2021,kumar_effect_2023,zhou_analytical_2015,starikov_angular-dependent_2022,shim_prediction_2013,angelo_trapping_1995,apostol_angular-dependent_2010,ko_atomistic_2011,zhou_bond-order_2018,hale_atomistic_2013}.
The major challenge is to achieve accurate energies and forces with respect to the quantum-mechanical calculations~\cite{voter_interatomic_1996,deringer_machine_2019}. The other class consists of machine-learning interatomic potentials (MLIPs)~\cite{behler_generalized_2007, bartok_gaussian_2010, thompson_spectral_2015, shapeev_moment_2016, drautz_atomic_2019, mishin_machine-learning_2021, batatia_mace_2023, xie_ultra-fast_2023} with complementary characteristics. MLIPs typically have lower transferability but higher accuracy. Indeed, current MLIPs demonstrate dramatically higher accuracy in predicting energies and forces compared to semi-empirical interatomic potentials and have also proven their effectiveness in predicting a wide range of derived properties~\cite{wang_machine_2024}.
Recently, MLIPs were also trained for the study of hydrogen diffusion and the energetics of hydrogenated unary and binary metallic systems~\cite{miwa_path_2019,qi_robust_2024,kwon_accurate_2023,yu_hydrogen_2024,angeletti_hydrogen_2024}.
However, these studies placed little emphasis on identifying
phase transitions during hydrogenation.
Moreover, the previous studies~\cite{miwa_path_2019,qi_robust_2024,kwon_accurate_2023,yu_hydrogen_2024,angeletti_hydrogen_2024} did not extensively examine hydrogen occupancies at the interstices, despite their critical role in phase transitions and in the interpretation of pressure--composition--isotherm (PCT) diagrams.
To develop MLIPs capable of addressing these aspects, their training sets need to be properly chosen to ensure accurate reproduction of the key features such as hydrogen ordering at low temperatures.

The present study provides a comprehensive investigation of hydrogen absorption in \ce{TiCr2H_x} in the C15 cubic and the C14 hexagonal Laves phases for a wide range of hydrogen concentrations of $0 < x \le 6$ based on DFT and MLIPs.
The MLIPs are trained through a series of active-learning schemes involving careful selection of training datasets to capture both low and high hydrogen concentrations.
The low-hydrogen configurations are systematically enumerated, and
the binding and the repulsion energies of hydrogen atoms in these configurations are analyzed based on DFT.
The basin-hopping Monte Carlo (BHMC) method is employed in conjunction with the MLIPs to identify minimum-energy hydrogen configurations as a function of hydrogen concentration and thus to locate phase transformations at \qty{0}{K}.
The thus obtained configurations of \ce{TiCr2H_x} and their formation-enthalpy profiles are analyzed to understand the phase transformations and detailed hydrogen distributions among the interstices.

\section{Methodology}
\label{sec:Methodology}

\subsection{Crystal structures and interstices}
\label{sec:structures}

\begin{table}[!tb]
\centering
\scriptsize
\caption{Fractional coordinates of the metal atoms (A=Ti and B=Cr) and geometric centers of the interstices (B\textsubscript{4}, AB\textsubscript{3}, and A\textsubscript{2}B\textsubscript{2}) for both the C15 cubic \ce{AB2} Laves phase with the space group $Fd\bar3m$ (227) (with the origin choice 2, i.e., the origin at the inversion center) and the C14 hexagonal \ce{AB2} Laves phase with the space group $P6_3/mmc$ (194).
The ``Wyckoff'' column provides the Wyckoff letter and the multiplicity for each position.
The values in bold are constrained by symmetry, while the others refer to the ideal case of a close-packed hard-sphere model with an atomic-radius ratio of $r_\mathrm{A}/r_\mathrm{B} = (3/2)^{1/2} \approx 1.225$ \cite{thoma_geometric_1995} and a $c/a$ ratio of $(8/3)^{1/2} \approx 1.633$ for the C14 phase.
The subscripts at the Wyckoff letters distinguish the interstices and follow the notation of Shoemaker and Shoemaker~\cite{shoemaker_concerning_1979}.}
\label{tab:positions_ideal}
\begin{tabular}{ccc*3{C{12mm}}}
\toprule
    & Type & Wyckoff & $x$ & $y$ & $z$ \\
\midrule
C15
& \ce{A}    & $8a$  & $\mathbf{1/8}$ & $\mathbf{1/8}$ & $\mathbf{1/8}$ \\
& \ce{B}    & $16d$ & $\mathbf{1/2}$ & $\mathbf{1/2}$ & $\mathbf{1/2}$ \\
\cmidrule{2-6}
& \ce{B4}   & $8b$  & $\mathbf{3/8}$ & $\mathbf{3/8}$ & $\mathbf{3/8}$ \\
\cmidrule{2-6}
& \ce{AB3}  & $32e$ & $9/32$ & $9/32$ & $9/32$ \\
\cmidrule{2-6}
& \ce{A2B2} & $96g$ & $5/16$ & $5/16$ & $1/8$ \\
\midrule
C14 & \ce{A}    &  $4f$  & $\mathbf{1/3}$ & $\mathbf{2/3}$ & $1/16$ \\
    & \ce{B}    &  $2a$  & $\mathbf{0}$ & $\mathbf{0}$ & $\mathbf{0}$ \\
    &           &  $6h$  & $-1/6$ & $-2/6$ & $\mathbf{1/4}$ \\
\cmidrule{2-6}
    & \ce{B4}   & $4e$   & $\mathbf{0}$ & $\mathbf{0}$ & $3/16$ \\
\cmidrule{2-6}
    & \ce{AB3}  & $4f$   & $\mathbf{1/3}$ & $\mathbf{2/3}$ & $43/64$ \\
    &           & $12k_1$& $1/8$ & $1/4$ & $23/64$ \\
\cmidrule{2-6}
    & \ce{A2B2}
      & $6h_1$  & $ 5/24$ & $ 5/12$ & $\mathbf{1/4}$ \\
    & & $6h_2$  & $11/24$ & $11/12$ & $\mathbf{1/4}$ \\
    & & $12k_2$ & $13/24$ & $13/12$ & $1/8$ \\
    & & $24l$   & $ 1/24$ & $ 1/ 3$ & $9/16$ \\
\bottomrule
\end{tabular}
\end{table}

\begin{figure}[!tb]
    \centering
    \includegraphics{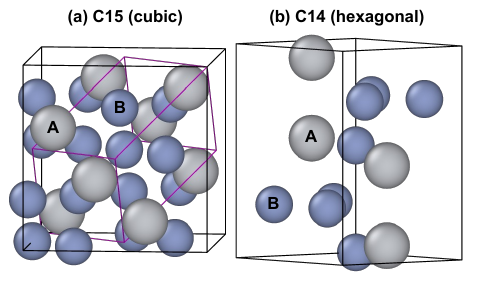}
    \caption{Crystal structures of (a) the C15 cubic and (b) the C14 hexagonal Laves phases.
    Elements A (Ti) and B (Cr) are shown as gray and blue spheres, respectively.
    The conventional unit cell of the C15 cubic phase is emphasized by the black lines, and the primitive unit cell is indicated by the purple lines.
    Visualization is performed using OVITO~\cite{stukowski_visualization_2009}.}
    \label{fig:CS}
\end{figure}

Laves phases consist of two metal elements, A and B, with a composition of \ce{AB2}.
The C15 Laves phase belongs to the space group $Fd\bar{3}m$ (227), 
while the C14 Laves phase belongs to the space group $P6_3/mmc$ (194).
The metal-atom positions of the C15 and the C14 Laves phases are shown in Table~\ref{tab:positions_ideal}.
The crystal structures of the C15 and the C14 phases are visualized in Fig.~\ref{fig:CS}.
In experiments for \ce{TiCr2},
the C15 cubic phase is found at low temperatures and the C14 hexagonal phase at high temperatures~\cite{sluiter_phase_1991,mebed_computer_1998}.
Hydrogen atoms may be accommodated in the tetrahedral interstices of the \ce{AB2} Laves phases.
These interstices are classified into three types based on the surrounding metal atoms: B$_4$, AB$_3$, and A$_2$B$_2$, and they are further subdivided by symmetry;
the C15 cubic and the C14 hexagonal Laves phases have three and seven symmetrically inequivalent interstices, respectively.
Table~\ref{tab:positions_ideal} also provides the positions of these interstices, along with their local chemical environments and their site symmetries.

\subsection{Ab initio calculations}
\label{sec:methods:DFT}

Density functional theory (DFT) calculations were conducted using VASP~\cite{furthmuller_dimer_1996,kresse_efficiency_1996,kresse_ultrasoft_1999}
with plane-wave basis sets and the projector augmented-wave (PAW) method~\cite{blochl_projector_1994}.
The exchange--correlation energy was obtained using the generalized gradient approximation (GGA) in the Perdew--Burke--Ernzerhof (PBE) form~\cite{perdew_generalized_1996}.
The valence electron configurations in the PAW potentials were $[\text{Ne}] 3s^2 3p^6 4s^2 3d^2$, $[\text{Ar}] 4s^1 3d^5$ and $1s^1$ for Ti, Cr, and H atoms, respectively.
The energy cutoff was set to 400\,eV.
Unless otherwise specified, all calculations were conducted with supercell models of a $2 \times 2 \times 2$ expansion of the primitive C15 unit cell and a $2 \times 2 \times 1$ expansion of the C14 unit cell, each containing 48 metal atoms.
The reciprocal space was sampled with a $\Gamma$-centered k-point mesh of $4 \times 4 \times 4$ for both the C15 and the C14 supercell models along with the Methfessel--Paxton smearing \cite{methfessel_high-precision_1989} with a width of 0.1\,eV.
Electronic self-consistent-field iterations were performed until the energy convergence threshold of $1 \times 10^{-8}$\,eV was reached.
Ionic relaxation was conducted using the conjugate gradient method until forces were below $1 \times 10^{-3}$\,eV/{\AA}.
Spin-polarized and non-spin-polarized calculations were initially compared. The results revealed that across all hydrogen concentrations in the database, magnetic moments do not influence predictions of energies and forces on atoms.
Furthermore, experimental studies have shown that \ce{TiCr2} Laves phases are paramagnetic~\cite{hiebl_proton_1982,abel_magnetic_1968}.
We thus concluded that non-spin-polarized calculations are sufficient to accurately model hydrogen absorption properties in the \ce{TiCr2} Laves phases and used them throughout the present study.

\subsection{Moment tensor potentials}
\label{sec:methods:MTP}


Moment tensor potentials (MTPs)~\cite{shapeev_moment_2016,podryabinkin_active_2017,novikov_mlip_2020,podryabinkin_mlip-3_2023} are a class of MLIPs widely used for various material systems~\cite{novikov_magnetic_2022,jung_high-accuracy_2023,ou_atomistic_2024,zhang_ab_2025,forslund_ab_2021,zhu_fully_2021,jung_dynamically_2023,zotov_moment_2024,srinivasan_anharmonicity_2023,
gubaev_finite-temperature_2021,
zhou_thermodynamics_2022,
xu_strong_2023,
luo_set_2023,
forslund_thermodynamic_2023,
gubaev_performance_2023,
srinivasan_electronic_2024,
xu_accurate_2024,
xu_origin_2024,
zhu_accelerating_2024,
zhu_melting_2024,
dash_recent_2022}.
In the present work, we developed MTPs to investigate the energetics of hydrogenated \ce{TiCr2H_x} in the C15 cubic and the C14 hexagonal Laves phases for a wide concentration range of $0 < x \le 6$.
The MLIP-2 software~\cite{novikov_mlip_2020} was employed for the training.
The MTPs were initially trained on dilute hydrogen concentrations---configurations containing no more than one hydrogen atom---with testing their accuracy as a function of the number of MTP parameters, i.e., the MTP complexity level~\cite{novikov_mlip_2020}.
As detailed in Sec.~\ref{sec:results:MTP_level}, the MTP level of 16 achieved the required accuracy and was thus considered in subsequent training steps for the wide range of hydrogen configurations.
The loss function to be minimized was defined as the sum of the errors in energy, forces on atoms, and stress, scaled such that every atom from every configuration in the training set contributes equally.
The weights in the loss function for energy (eV), forces (eV/\AA), and stress times volume (eV) were set to 1, 0.01, and 0.001, respectively.
The MTP parameters were initialized with random values and then optimized at each step when the training set was updated by inheriting the values optimized in the previous step. 
The optimization was done using the Broyden--Fletcher--Goldfarb--Shanno method as implemented in the MLIP-2 software.

Note that, in the present study, we are mainly interested in the formation enthalpy of the hydrogenated \ce{TiCr2H_x} in each of the C15 and the C14 phases rather than the energy differences between the two phases.
Therefore, the MTPs were trained for each phase individually so that a higher accuracy on the formation enthalpy of hydrogenation could be achieved for each phase.
When necessary, the MTPs for the C15 and the C14 phases are explicitly referred to as the C15-MTP and the C14-MTP, respectively.

\subsection{Training datasets}
\label{subsec:Data-sets}

The training datasets for the MTPs were constructed in multiple steps detailed below, with each step introducing different types of hydrogen configurations and retraining the MTPs.
The resulting MTPs become sufficiently transferable
for various general hydrogen configurations visited during the Monte Carlo simulations detailed in Sec.~\ref{sec:methods:BHMC}.
This allows us to investigate the formation enthalpy of~\ce{TiCr2H_x} with near-DFT accuracy, accelerated by the MTPs, and thus to obtain the minimum-energy hydrogen configurations as a function of hydrogen concentration.

In later steps, we employed an active-learning approach~\cite{podryabinkin_active_2017,
gubaev_accelerating_2019},
where candidate configurations were filtered according to their extrapolation grades defined based on D-optimality, before performing DFT calculations for these and incorporating them to the current training dataset.
The active-learning approach was used in many previous studies to accelerate the training of MTPs~\cite{
gubaev_finite-temperature_2021,
gubaev_performance_2023,
bock_active_2024,
erhard_modelling_2024,
mismetti_automated_2024,
ou_atomistic_2024,
xu_origin_2024}.
The present training scheme applies the active learning repeatedly for various hydrogen configurations among the interstices in \ce{TiCr2} Laves phases to enhance the transferability of MTPs.
A conceptually similar approach was applied by Xu~\textit{et al.}~\cite{xu_origin_2024} for \ce{Ni3Al}, where various types of structural defects were incorporated stepwise.
A brief overlook of the MTP-training flowcharts are presented in Fig.~\ref{fig:main}.
Sec.~\textcolor{black}{S1} in the Supplementary Materials (SM) presents a more detailed flowchart at each step.

\begin{figure}[!tb]
    \centering
    \includegraphics[width=\linewidth]{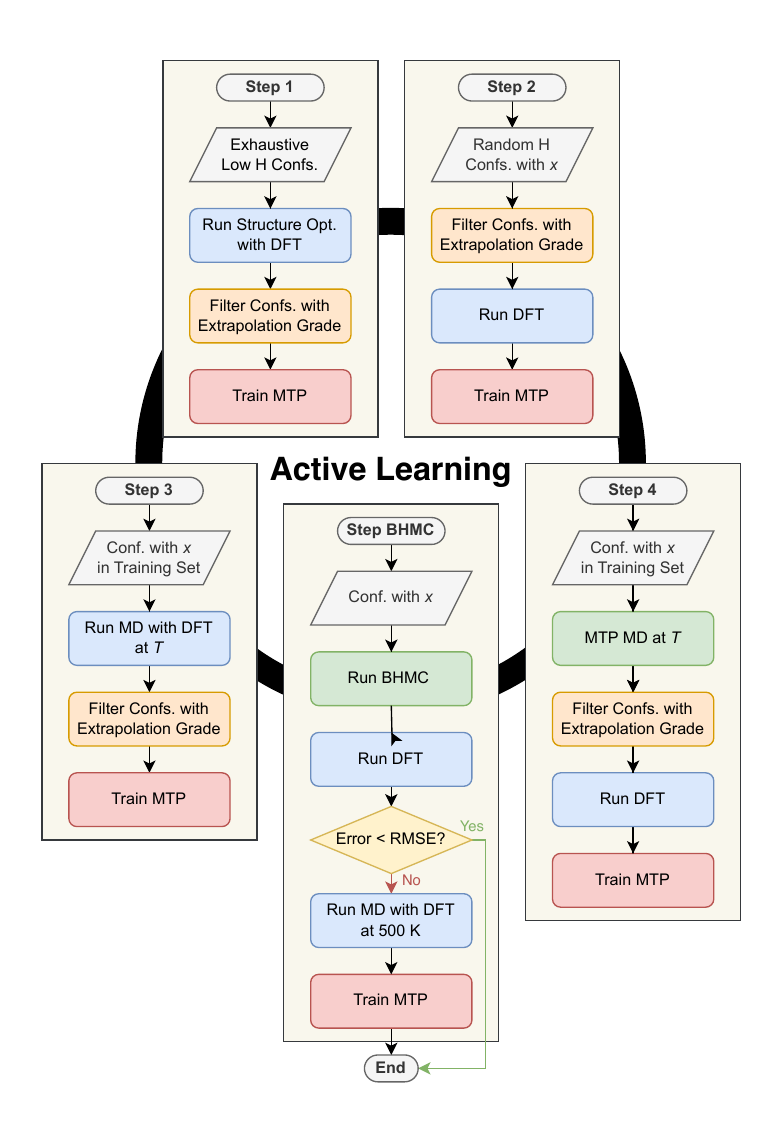}
    \caption{Flowcharts depicting the methods to generate various types of configurations used for the training datasets of the MTPs.
    The technique also includes active-learning parts employing the already trained MTPs to look for additional configurations that further enhance the transferability of the MTPs.}
    \label{fig:main}
\end{figure}

\subsubsection{Step 1: Low hydrogen concentrations}
\label{sec:step-0}
The first step focuses on low hydrogen concentrations.
Specifically, \ce{TiCr2} without hydrogen atoms as well as one up to six hydrogen atoms within the 48-metal-atom cells were considered.
The configurations with hydrogen atoms were generated by enumerating all symmetrically inequivalent local hydrogen configurations exhaustively within certain criteria.
Thus, we sampled unique local hydrogen configurations representing the interaction between hydrogen atoms close to each other, which may be an important factor for hydrogen capacity and ordering in the alloy.
The detailed procedure is given in~\ref{sec:graph_theory}, where graph theory is employed to describe the connection of the interstices.
Graph distances among the hydrogen atoms up to four and three were considered for the configurations with two and more hydrogen atoms, respectively.
A total of 1019 and 3766 configurations were obtained for the C15 cubic and the C14 hexagonal Laves phases, respectively.
For each configuration, hydrogen atoms were initially placed at the geometric centers of the interstices, and then structure relaxation was performed  in DFT.
The relaxation involved cell parameters together with atomic positions to incorporate the volume expansion due to the hydrogen atoms.
All the ionic steps from all configurations traversed during the relaxation were once accumulated. 
The training dataset was then initialized by sub-sampling \qty{10}{\percent} configurations in the accumulated ones, and the initial MTP was trained based on them.
Then, the extrapolation grades were calculated for the remaining configurations based on the initial training dataset, and the configurations showing the extrapolation grades $\gamma \gtrsim 1$, i.e., substantially extrapolative, were appended to the training dataset, followed by the retraining of the MTP.
This procedure yielded the first-generation training datasets and the corresponding MTPs as the output.

\subsubsection{Step 2: Random hydrogen configurations}
\label{sec:step-1}

In the next step, we extended the training dataset to include a wider range of hydrogen concentrations, specifically $0 < x \le 6$ for \ce{TiCr2H_x}, with a large number of randomly generated configurations.
A straightforward approach might be just adding all the generated configurations to the training dataset.
However, this approach is neither efficient nor effective, as many of the generated configurations could be similar to each other, thus failing to significantly extend the configurational space covered by the MTP.
Additionally, performing DFT calculations on all these configurations, including the redundant configurations, is computationally expensive.
We therefore pre-filtered the configurations that can genuinely extend the existing configurational space among the generated configurations based on the extrapolation grade, which reduced the number of DFT simulations needed.
We considered hydrogen concentrations in the range of $0 < x \leq 3$ with a step of 0.25 starting from $x = 0.25$ (4 hydrogen atoms in the 48 metal-atom cells), and also $x = 6$ (96 hydrogen atoms).
The step started from the lowest concentration ($x = 0.25$).
For each concentration, more than \num{25000} configurations were generated, in each of which hydrogen atoms were randomly distributed among the interstices.
Next, in each configuration, the positions of atoms were perturbed according to a normal distribution with the standard deviation of \qty{0.1}{\angstrom}.
Then, the extrapolation grades were calculated for these configurations based on the current training set, and only the configurations showing the extrapolation grades $\gamma \gtrsim 1$ were selected.
After this filtering, the selected configurations were computed with DFT and then added to the present training dataset.
The MTP was then retrained based on the extended dataset starting from the previous parameters.
Once this process was completed, we moved to the next hydrogen concentration, and the procedure was repeated.

\subsubsection{Step 3: DFT MD}
\label{sec:step-2}

In this step, we extended the training dataset with the configurations from \textit{ab initio} MD trajectories.
These thermally driven configurations are expected to fine-tune the MTP parameters to make the potential energy surface smoother.

We considered hydrogen concentrations with a step of 0.5 for ranges of $0 < x \le 3$ for the C15 phase and $0 < x \le 3.5$ for the C14 phase.
The step started from the lowest concentration.
For each concentration, one configuration was selected randomly from the present training dataset.
From the selected configuration, 11 configurations were made by applying volumetric compression or expansion by up to \qty{15}{\percent} to cover a wider region of the configurational space and to thus enhance the stability of the trained MTPs.
For each of these configurations, \textit{ab initio} MD simulation was performed in the \textit{NVT} ensemble with an Nosé--Hoover thermostat~\cite{nose_unified_1984,hoover_canonical_1985} at a temperature of \qty{500}{K} using a time step of \qty{1}{fs}.
The MD simulations ran for \num{100} steps, with the trajectories being saved.
The trajectories were further filtered using the extrapolation grade based on the current training database, similar to Step 2 (Sec.~\ref{sec:step-1}).
Configurations with an extrapolation grade greater than 1.1 were then added to the current training dataset, and new MTPs were retrained.
Once this process was completed, we moved to the next hydrogen concentration, and the procedure was repeated.

\subsubsection{Step 4: MTP MD}
\label{sec:step-3}

The MTPs trained up to the above steps already cover a wide range of hydrogen concentrations.
To retune the MTPs again for \ce{TiCr2H_x} at low-hydrogen-concentration range, we again extended the training dataset based on MD simulations using the present MTP along with the active-learning scheme.

We considered two hydrogen concentrations, one without hydrogen and one with one hydrogen atom in the simulation cell.
We first picked up one configuration without hydrogen from the present training dataset.
We then ran an MD simulation for this configuration under the \textit{NVT} ensemble with a Nosé--Hoover-chain thermostat \cite{martyna_nosehoover_1992}, with a time step of \qty{5}{fs}, a temperature damping factor of \qty{0.5}{ps}, and a chain length of 3, implemented in LAMMPS~\cite{thompson_lammps_2022} together with the MLIP-2 interface~\cite{novikov_mlip_2020}.
Within the MD trajectory, all the configurations with extrapolation grades $\gamma \gtrsim 1$ were saved.
Once the MD trajectory reaches an extrapolation grade greater than 10, the simulation is terminated, as such a high grade often leads to a breakdown of the subsequent MD due to inaccurate atomic forces predicted by the current MTP.
Next, single-point DFT calculations were performed on the saved extrapolative configurations.
The MTP was then retrained by incorporating these configurations into the current training set.
This process was repeated at 300\,K, 500\,K and 700\,K until the MD trajectory becomes fully interpolative, and thus the MD simulation can run until the end of the given MD simulation time, i.e., \qty{0.5}{ns} (\num{100000} MD steps).
The procedure was repeated then for the concentration with one hydrogen atom.
Notably, the thus trained MTPs were found to be computationally stable and did not break MD runs anymore not only for the dilute but also for higher hydrogen concentrations.
This implies that the MTPs are still robust for the whole hydrogen-concentration range considered in the present study.

\subsection{Minimum-energy hydrogen configurations}
\label{sec:methods:BHMC}

We utilize our trained MTPs to identify the minimum-energy configurations of hydrogen atoms in \ce{TiCr2} at given hydrogen concentrations.
These minimum-energy configurations provide valuable insights into the phase transition during hydrogenation and distribution of hydrogen atoms within these phases.
In this section, we present an extensive search method for identifying the minimum-energy configurations across varying hydrogen concentrations using our trained MTPs.
The challenge for the present systems is that, apart from dynamically unstable configurations (cf.~Sec.~\ref{sec:results:repulsion_faces}), every hydrogen configuration has a corresponding local minimum on the potential-energy surface.
Consequently, a purely gradient-based local minimization method is insufficient for locating the global minimum on the potential-energy surface.
To address this, we employ a global optimization approach based on the canonical Metropolis MC sampling with a swap of occupied and unoccupied interstices, followed by a gradient-based local minimization method.
This is essentially aligned with the basin-hopping MC (BHMC) approach~\cite{wales_global_1997}, also known as the Monte Carlo minimization method \cite{li_monte_1987}, and hence we also refer to our approach as the BHMC method.
After each swap, the positions of both metal and hydrogen atoms are relaxed with fixing the cell parameters, reaching a local-minimum on the potential-energy surface.
The obtained configuration is accepted or rejected based on the Boltzmann factor for the energies of the local minima of the present and the previously-accepted steps with a given temperature.
The BHMC simulations were performed at multiple constant temperatures: \qtylist{0;50;100;200;300}{\kelvin}.
A finite temperature was essential for facilitating multiple swaps, allowing the simulation to explore different configurations until a local minimum with a lower energy was identified.
Each BHMC simulation included at least \num{30000} swaps per 48-metal-atom simulation cell.
Notably, the emergence of a fully ordered hydride structure required a significantly larger number of steps---up to one million.

Once the BHMC simulation is complete, the minimum-energy configuration encountered during the simulation is selected.
(Note that this is not necessarily the final configuration in the BHMC simulation.)
The selected configuration is then further relaxed, allowing now also cell relaxation in addition to the atomic positions.
The BHMC simulation is then restarted using the newly optimized cell parameters.
This iterative process continues until no further energy minimization is observed, indicating that the minimum-energy configuration at the given hydrogen concentration has been reached.
Separating cell-parameter relaxation from the BHMC simulation accelerated convergence of the local-relaxation part in the present systems.

To facilitate the identification of minimum-energy configurations, various initial hydrogen configurations were tested. These included configurations in which one or more sets of symmetrically distinct interstices were completely filled. Additionally, some configurations were derived from the minimum-energy configuration of a system with one more or one fewer hydrogen atoms than the target concentration, which helps smooth the formation-enthalpy profile.
BHMC simulations were also performed on specific sets of symmetrically distinct interstices.
These constrained simulations accelerated the convergence to minimum-energy configurations, particularly for ordered structures.

The identified minimum-energy configuration is used also for further retraining the MTP.
Specifically, the energy of the configuration is evaluated using DFT, and the DFT energy is compared with the prediction from the current MTP.
If the MTP-predicted energy differs from the DFT value more than the RMSE of the MTP for the current training set, \textit{ab initio} MD simulation is performed for the minimum-energy configuration at \qty{500}{K} with otherwise the same setting as Sec.~\ref{sec:step-2}, and the configurations in the trajectories are appended to the current training dataset.
The MTP is then retrained with the updated training set, and the BHMC simulations are rerun using the retrained MTP.
This process continues until the energy error of the minimum-energy configuration remains within the RMSE for the training dataset.
\section{Results and Discussion}
\label{sec:Results}

\subsection{Binding energies of single hydrogen atoms}
\label{sec:results:binding_single_H}

We first discuss the binding energy of a single hydrogen atom in the \ce{TiCr2} Laves phases.
The enthalpy of formation of the hydrogenated \ce{TiCr2H_x} Laves phase per formula unit, $\Delta_\mathrm{f} H$, is given by
\begin{equation}
    \label{eq:sol}
     \Delta_\mathrm{f}H = E\left(\ce{TiCr2H_x}\right)-E\left(\ce{TiCr2}\right)-x\cdot\dfrac{E(\ce{H2})}{2},
\end{equation}
where $E\left(\ce{TiCr2H_x}\right)$ and $E\left(\ce{TiCr2}\right)$ are the energies of hydrogenated \ce{TiCr2H_x} and non-hydrogenated \ce{TiCr2} per formula unit after the relaxation of both atomic positions and cell parameters, and $E(\ce{H2})$ is the energy of an isolated hydrogen molecule.
The energy of an isolated \ce{H_2} molecule was calculated by placing it in a cubic box with a side length of \qty{10}{\angstrom}, and atomic relaxation was performed using $\Gamma$-point sampling. This calculation resulted in a bond length of \qty{0.7505}{\angstrom}, in reasonable agreement with experiments~(\qty{0.741 44}{\angstrom}~\cite{huber_constants_1979}).
The binding energy per hydrogen atom $E_\mathrm{b}$ is obtained by normalizing $\Delta_\mathrm{f} H$ by the number of hydrogen atoms;
\begin{align}
E_\mathrm{b} = \frac{1}{n_\mathrm{H}} \cdot \Delta_\mathrm{f}H,
\end{align}
where $n_\mathrm{H}$ is the number of hydrogen atoms per formula unit.
A more negative $E_\mathrm{b}$ value indicates higher hydrogen solubility at the corresponding sites.

Table~\ref{tab:solution_energy_single} presents the hydrogen binding energies at symmetrically distinct interstitial sites, as obtained from the DFT simulations.
In both the C15 cubic and the C14 hexagonal Laves phases, the \ce{A2B2} interstices are the most energetically favorable for hydrogen.
The \ce{AB3} and the \ce{B4} interstices are much less favorable, and the binding at the \ce{B4} interstices is even highly endothermic.
Several computational and experimental studies have also shown that the hydrogen solubility is the highest at the \ce{A2B2} interstitial sites in \ce{TiCr2}~\cite{li_hydrogen_2009,li_achieving_2019} and other Laves-phase alloys~\cite{%
gesari_hydrogen_2010,
merlino_dft_2016,
jiang_influence_2022,
yartys_laves_2022},  
consistent with the present findings.
This may be because the \ce{A2B2} interstitial sites have the largest available volumes, followed by the \ce{AB3} and the \ce{B4} sites (see, e.g., Table~1 in Ref.~\cite{mohammadi_high-entropy_2022}).

\begin{table}[tb]
\centering
\scriptsize
\let\TPToverlap=\TPTrlap
\caption{Binding energies $E_\mathrm{b}$ (eV/H) of a single hydrogen atom at the interstices of the C15 cubic and the C14 hexagonal \ce{TiCr2} Laves phases, as obtained from the DFT simulations.
For comparison, values from previous DFT-based simulations for the C15 phase~\cite{li_hydrogen_2009} and the C14 phase~\cite{jiang_influence_2022} are also shown.\tnote{1,2,3\textcolor{black}{,4}}}
\label{tab:solution_energy_single}
\begin{threeparttable}
\begin{tabular}{
ccc
S[table-format=7.7,retain-explicit-plus]
S[table-format=7.7,retain-explicit-plus]
}
\toprule
& Type & Site & \multicolumn{1}{c}{Present} & \multicolumn{1}{c}{Others~\cite{li_hydrogen_2009,jiang_influence_2022}} \\
\midrule
C15                  & \ce{B4}   & $8b$    & +0.599 & +1.68 \\
\cmidrule{2-5}
                     & \ce{AB3}  & $32e$   & -0.016 & -0.02 \\
\cmidrule{2-5}
                     & \ce{A2B2} & $96g$   & -0.217 & -0.22 \\
\midrule
C14                  & \ce{B4}   & $4e$    & +0.566 & +0.6576  \\
\cmidrule{2-5}
                     & \ce{AB3}  & $4f$    & +0.077 & +0.1575  \\
                     &           & $12k_1$ & -0.019 & +0.0404  \\
\cmidrule{2-5}
                     & \ce{A2B2} & $6h_1$  & -0.170 & -0.1028 \\
                     &           & $6h_2$  & -0.249 & -0.1881 \\
                     &           & $12k_2$ & -0.173 & -0.1077 \\
                     &           & $24l$   & -0.185 & -0.1405 \\
\bottomrule
\end{tabular}
\footnotesize
\begin{tablenotes}
\item [1] From Li \textit{et al.}~\cite{li_hydrogen_2009}, the values without zero-point energies have been taken to be consistent with the other results in the present table.
\item [2] \textcolor{black}{The significantly higher binding energy obtained by Li~\textit{et al.}~\cite{li_hydrogen_2009} particularly for the C15 \ce{B4} site is probably due to their boundary conditions where both the cell parameters and the atomic coordinates were fixed. See details in Sec.~S4.1 in the SM.}
\item [3] Fig.\,S3 in Jiang \textit{et al.}~\cite{jiang_influence_2022} reveals that their $6h_1$ and $6h_2$ sites are reversed compared to those in Shoemaker and Shoemaker~\cite{shoemaker_concerning_1979} and thus to those in the present manuscript.
For the present table, they have been reversed back to be consistent with our convention.
\item [4] Jiang~\textit{et al.}~\cite{jiang_influence_2022} show values substantially less negative than our values, which is likely due to different computational settings. For example, they used the DFT+\textit{U} method.
\end{tablenotes}
\end{threeparttable}
\end{table}

For the C14 hexagonal phase, the binding energies differ substantially even among the \ce{A2B2} interstices.
Specifically, the $6h_2$ interstices are the most energetically favorable for hydrogen ($E_\mathrm{b}$ = \qty[number-color=red]{-249}{meV/H}), followed by the $24l$ interstices (\qty[number-color=red]{-185}{meV/H}).
The binding energies at the $12k_2$ interstices (\qty[number-color=red]{-170}{meV/H}) and the $6h_1$ interstices (\qty[number-color=red]{-173}{meV/H}) are further less negative, and they differ only by \qty{3}{meV/H}.
Overall, hydrogen solubility at the seven distinct interstices is ordered as $6h_2 > 24l > 12k_2 {~\color{red}\approx~} 6h_1 > 12k_1 > 4f > 4e$, in qualitative agreement with previous DFT+\textit{U} simulations~\cite{jiang_influence_2022}.
The substantial differences in energetic favorability influence hydrogen occupancy at these interstices, particularly  at low hydrogen concentrations, as confirmed in Sec.~\ref{sec:results:C14-0-1} and \ref{sec:results:C14-0-6}.

\color{black}
When the zero-point energies (ZPEs) of the hydrogen atoms are taken into account (Sec.~S4.2 in the SM), the binding energies increase by \qtyrange[range-phrase=--,range-units=single]{180}{190}{meV/H}, \qtyrange[range-phrase=--,range-units=single]{140}{150}{meV/H}, and \qtyrange[range-phrase=--,range-units=single]{110}{120}{meV/H} for the \ce{B4}, the \ce{AB3}, and the \ce{A2B2} sites, respectively.
However, the relative sequence of the binding energies among the interstitial sites in each Laves phase is almost unchanged.
Therefore, the ZPE correction would not affect the qualitative conclusions of the subsequent analyses below and is omitted for the sake of feasibility.
\color{black}

\subsection{Short-range hydrogen--hydrogen repulsion}

Hydrogen absorption properties of hydrogen-storage alloys should be strongly influenced by the distribution of hydrogen atoms among interstices.
To gain deeper insight, we systematically analyze hydrogen--hydrogen repulsion in the \ce{TiCr2} Laves phases.
The repulsion energy per hydrogen pair is given by
\begin{align}
E_\mathrm{rep} = E_\mathrm{b}^{S_1+S_2} - E_\mathrm{b}^{S_1} - E_\mathrm{b}^{S_2},
\label{eq:rep}
\end{align}
where $E_\mathrm{b}^{S_1}$ and $E_\mathrm{b}^{S_2}$ represent the binding energies of isolated hydrogen atoms occupying sites $S_1$ and $S_2$, respectively, while $E_\mathrm{b}^{S_1+S_2}$ denotes the cumulative binding energy when both the sites are occupied simultaneously.

\subsubsection{Face-sharing interstices}
\label{sec:results:repulsion_faces}

\begin{table}[tb]
\centering
\scriptsize
\caption{Cumulative hydrogen binding energies (eV/2H) and hydrogen repulsion energies (meV/pair) of all symmetrically distinct face-sharing configurations in the \ce{TiCr2} Laves phases obtained using DFT.
Dynamically unstable configurations are indicated as ``unstable''.}
\label{tab:repulsion_faces}
\begin{tabular}{ccccS[table-format=3.3]S[table-format=5.0]}
\toprule
&
Pair &
Type &
Face &
\multicolumn{1}{c}{$E_\mathrm{b}^{S_1+S_2}$} &
\multicolumn{1}{c}{$E_\mathrm{rep}$} \\
\midrule
C15
 & $8b$--$32e$      & \ce{B4}--\ce{AB3}    & \ce{B3}  & +0.750     & +167       \\
 & $32e$--$96g$     & \ce{AB3}--\ce{A2B2}  & \ce{AB2} & +0.243     & +475       \\
 & $96g$--$96g$     & \ce{A2B2}--\ce{A2B2} & \ce{AB2} & -0.035     & +399       \\
 & $96g$--$96g$     & \ce{A2B2}--\ce{A2B2} & \ce{A2B} & {unstable} & {unstable} \\
\midrule
C14
 & $4e$--$4e$       & \ce{B4}--\ce{B4}     & \ce{B3}  & +1.554     & +421       \\
 & $4e$--$12k_1$    & \ce{B4}--\ce{AB3}    & \ce{B3}  & +0.756     & +209       \\
 & $4f$--$4f$       & \ce{AB3}--\ce{AB3}   & \ce{B3}  & +0.262     & +107       \\
 & $4f$--$12k_2$    & \ce{AB3}--\ce{A2B2}  & \ce{AB2} & +0.295     & +391       \\
 & $12k_1$--$6h_1$  & \ce{AB3}--\ce{A2B2}  & \ce{AB2} & +0.295     & +484       \\
 & $12k_1$--$24l$   & \ce{AB3}--\ce{A2B2}  & \ce{AB2} & {unstable} & {unstable} \\
 & $6h_1$--$6h_2$   & \ce{A2B2}--\ce{A2B2} & \ce{A2B} & {unstable} & {unstable} \\
 & $6h_2$--$12k_2$  & \ce{A2B2}--\ce{A2B2} & \ce{AB2} & -0.006     & +416       \\
 & $12k_2$--$24l$   & \ce{A2B2}--\ce{A2B2} & \ce{A2B} & {unstable} & {unstable} \\
 & $24l$--$24l$     & \ce{A2B2}--\ce{A2B2} & \ce{AB2} & -0.012     & +359       \\
 & $24l$--$24l$     & \ce{A2B2}--\ce{A2B2} & \ce{A2B} & {unstable} & {unstable} \\
\bottomrule
\end{tabular}

\end{table}

\begin{figure}[tb]
    \centering
    \includegraphics{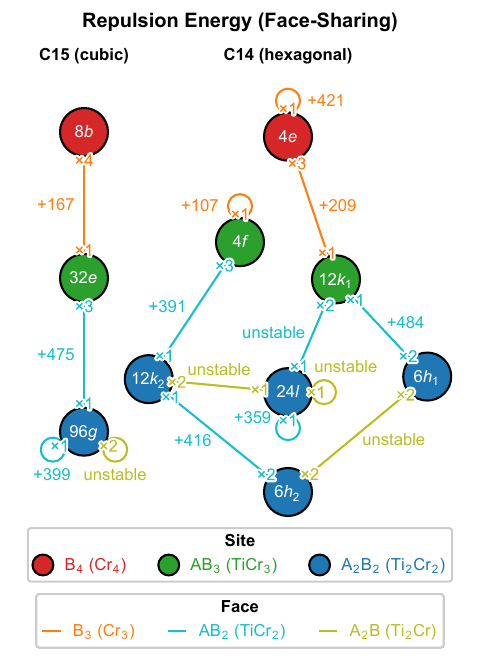}
    \caption{All symmetrically distinct face-sharing configurations together with their multiplicities and their hydrogen repulsion energies (meV/pair) in Table~\ref{tab:repulsion_faces}.}
    \label{fig:faces}
\end{figure}

\begin{figure*}[!tb]
\centering
\includegraphics[width=0.9\linewidth]{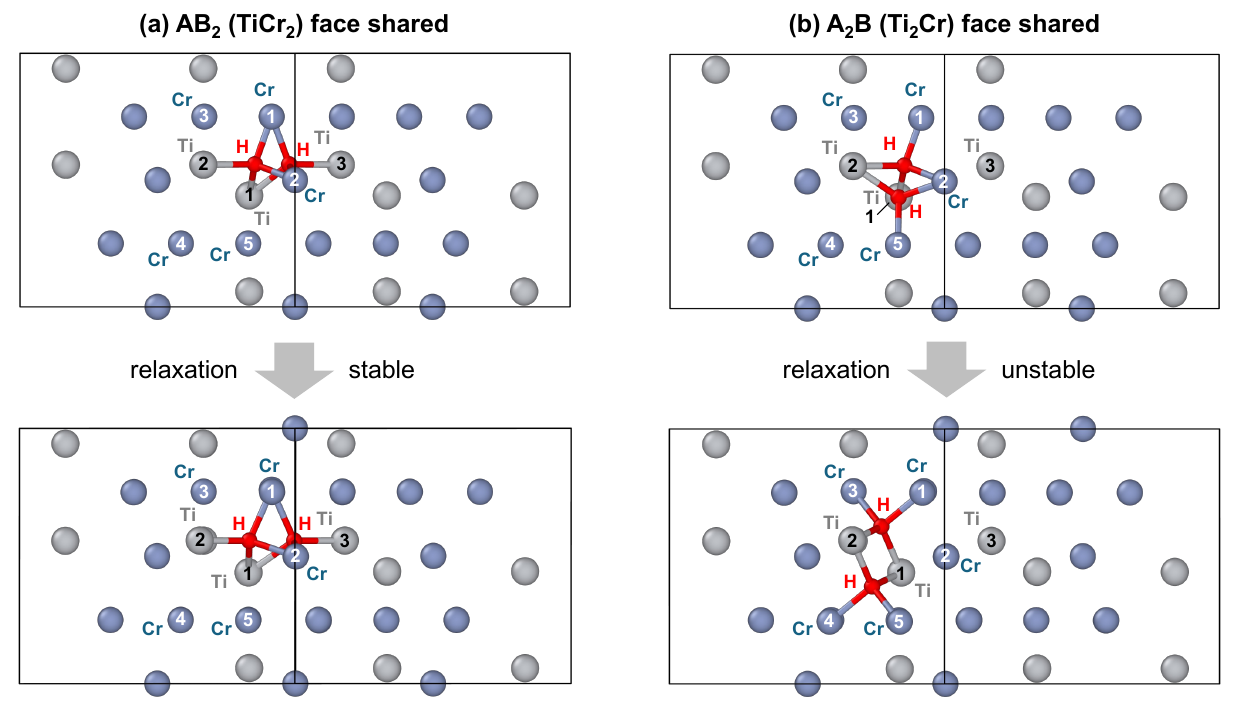}
\caption{Initial and relaxed configurations with two hydrogen atoms initially at face-sharing $24l$ tetrahedral interstices in C14 \ce{TiCr2}.
(a)~Configuration with a common \ce{AB2} (\ce{TiCr2}) face.
(b)~Configuration with a common \ce{A2B} (\ce{Ti2Cr}) face.
}
\label{fig:24l-24l}
\end{figure*}

Switendick proposed that distances between hydrogen atoms in metal hydrides remain greater than 2.1\,{\AA}~\cite{switendick_band_1979,rao_switendick_1985}.
Shoemaker and Shoemaker~\cite{shoemaker_concerning_1979} later proposed an exclusion rule for hydrogen atoms in the Laves phases, stating that face-sharing interstices cannot simultaneously accommodate hydrogen atoms.
They also found that the distances between the centers of face-sharing interstices are smaller than 1.6\,{\AA}, thereby demonstrating that this exclusion rule aligns with the Switendick criterion.
Moreover, previous computational studies~\cite{radakovic_hydrogen_2013,rabahi_energetics_2017} have consistently corroborated this exclusion rule by showing that hydrogen atoms do not occupy face-sharing interstices in their investigated Laves phases, reinforcing the theoretical foundation of the Shoemaker–Shoemaker exclusion rule.

Table~\ref{tab:repulsion_faces} presents the cumulative binding energies and the repulsion energies of all symmetrically distinct face-sharing configurations of two hydrogen atoms in the \ce{TiCr2} Laves phases obtained from DFT.
Additionally, Fig.~\ref{fig:faces} visually summarizes these face-sharing configurations together with the repulsion energies.
The C15 cubic and the C14 hexagonal Laves phases exhibit 4 and 11 symmetrically distinct face-sharing configurations, respectively.
Notably, both the $96g$--$96g$ pair in the C15 phase and the $24l$--$24l$ pair in the C14 phase each exhibit two symmetrically distinct configurations, one sharing an \ce{AB2} face and the other an \ce{A2B} face in common.

Some face-sharing configurations are intrinsically dynamically unstable, leading to spontaneous hydrogen-atom migration during structural relaxation. Specifically, in such cases, one or both hydrogen atoms are expelled from their initial interstitial sites.
We systematically tested various initial hydrogen positions and confirmed the invariance of the observed instability.
Notably, all the \ce{A2B2}--\ce{A2B2} configurations that share a common \ce{A2B} face exhibit dynamic instability, indicating strong unfavorability.
The only exception is the unstable $12k_1$--$24l$ configuration in the C14 phase with a common \ce{AB2} face.
The expelled hydrogen atoms migrate to unoccupied nearest interstitial sites while still maintaining edge-sharing between the newly occupied interstices.
For instance, in the $12k_1$--$24l$ configuration, the hydrogen atom initially positioned at the $12k_1$ site migrates into a neighboring $24l$ interstice, which shares a common edge with the $24l$ site occupied by the immobile hydrogen atom.

The remaining face-sharing configurations are found to be dynamically stable, i.e., the hydrogen atoms stay within their initial interstices even after structural relaxation.
This stability persists even after perturbing the atomic positions and subsequently re-relaxing the structure, confirming that these configurations are intrinsically robust.
However, despite their dynamical stability, these hydrogen pairs exhibit significant repulsion energies, ranging from a minimum of \qty[number-color=red]{+107}{meV/pair} to a maximum of \qty{+484}{meV/pair}.
These findings indicate that such face-sharing occupations are energetically unfavorable at dilute hydrogen concentrations, aligning with the assumption proposed by Shoemaker and Shoemaker~\cite{shoemaker_concerning_1979}.
Furthermore, the maximum hydrogen--hydrogen distance for these face-sharing configurations is 1.22\,{\AA}, further supporting the validity of the Switendick criterion~\cite{switendick_band_1979,rao_switendick_1985}.

Among the \ce{A2B2}--\ce{A2B2} face-sharing configurations, those with a common \ce{AB2} remains dynamically stable  (e.g., Fig.~\ref{fig:24l-24l}(a) illustrating the \ce{TiCr2}-face-sharing $24l$--$24l$ interstices in C14 \ce{TiCr2}), and only those with a common \ce{A2B} face exhibit dynamic instability (e.g., Fig.~\ref{fig:24l-24l}(b) illustrating the \ce{Ti2Cr}-face-sharing $24l$--$24l$ interstices).
The difference can be understood from the perspectives of geometry and screened Coulomb interaction.

In the present DFT simulations for \ce{TiCr2}, the \ce{A2B} faces have areas in the range of 3.5--3.6\,{\AA}\textsuperscript{2}, which are substantially larger than those of \ce{B3} faces (2.5--2.7\,{\AA}\textsuperscript{2}) and \ce{AB2} faces (3.0--3.2\,{\AA}\textsuperscript{2}).
On the other hand, Li~\textit{et al.}~\cite{li_hydrogen_2009} demonstrated that the repulsion between hydrogen atoms in C15 \ce{TiCr2} fitted well to the screened Coulomb potential.
Assuming that the screening of Coulomb interaction is weaker through a common face with a larger area due to less impact from the metal atoms on the common face, the repulsion between the hydrogen atoms sharing a common \ce{A2B} faces should be stronger than a common \ce{AB2} face, resulting in the dynamical instability of the former case.
Further, a closer examination of the structure-relaxation trajectories reveals that, in \ce{A2B}-face-sharing configurations, all
jumps of hydrogen atoms during the relaxation are through other unshared \ce{A2B} faces.
For example, two hydrogen atoms in Fig.~\ref{fig:24l-24l}(b), initially located at the \ce{Ti2Cr}-face-sharing $24l$--$24l$ interstices, migrate through the faces labeled Ti1--Ti2--Cr1 and Ti1--Ti2--Cr5.
This suggests that hydrogen atoms can jump more easily through the larger \ce{A2B} faces.
Since \ce{A2B} faces exist only in \ce{A2B2} interstices, this may explain why hydrogen migration is predominantly observed within \ce{A2B2} interstices.
Note that the above discussion does not apply to the \ce{AB2}-face-sharing $12k_1$–$24l$ interstices, which should be regarded as an exceptional case due to subtle atomic interactions.
\color{black}

\subsubsection{Edge-sharing interstices}
\label{sec:results:repulsion_edges}

\begin{table}[!tb]
\centering
\scriptsize
\caption{Cumulative hydrogen binding energies (eV/2H) and hydrogen repulsion energies (meV/pair) of all symmetrically distinct edge-sharing configurations in the \ce{TiCr2} Laves phases obtained using DFT. $D$ is the graph distance between the occupied interstices (\ref{sec:graph_theory}).}
\label{tab:repulsion_edges}
\begin{tabular}{cccccS[table-format=3.3]S[table-format=5.0]}
\toprule
&
$D$ &
Pair &
Type &
Edge &
\multicolumn{1}{c}{$E_\mathrm{b}^{S_1+S_2}$} &
\multicolumn{1}{c}{$E_\mathrm{rep}$} \\
\midrule
C15
 & 2 & $8b$--$96g$      & \ce{B4}--\ce{A2B2}   & \ce{B2}  & +0.359     &  -23       \\
 & 2 & $32e$--$32e$     & \ce{AB3}--\ce{AB3}   & \ce{B2}  & -0.038     &   -6       \\
 & 2 & $32e$--$96g$     & \ce{AB3}--\ce{A2B2}  & \ce{B2}  & -0.190     &  +42       \\
 & 2 & $32e$--$96g$     & \ce{AB3}--\ce{A2B2}  & \ce{AB}  & -0.122     & +111       \\
 & 2 & $96g$--$96g$     & \ce{A2B2}--\ce{A2B2} & \ce{AB}  & -0.372     &  +61       \\
 & 2 & $96g$--$96g$     & \ce{A2B2}--\ce{A2B2} & \ce{A2}  & -0.330     & +104       \\
 & 2 & $96g$--$96g$     & \ce{A2B2}--\ce{A2B2} & \ce{AB}  & -0.357     &  +76       \\
\cmidrule(){2-7}
 & 3 & $96g$--$96g$     & \ce{A2B2}--\ce{A2B2} & \ce{A2}  & -0.426     &   +7       \\
\midrule
C14
 & 2 & $4e$--$12k_1$    & \ce{B4}--\ce{AB3}    & \ce{B2}  & +0.540     &   -7       \\
 & 2 & $4e$--$6h_1$     & \ce{B4}--\ce{A2B2}   & \ce{B2}  & +0.396     &   -0       \\
 & 2 & $4e$--$24l$      & \ce{B4}--\ce{A2B2}   & \ce{B2}  & +0.366     &  -15       \\
 & 2 & $4f$--$6h_2$     & \ce{AB3}--\ce{A2B2}  & \ce{B2}  & -0.161     &  +11       \\
 & 2 & $4f$--$12k_2$    & \ce{AB3}--\ce{A2B2}  & \ce{B2}  & -0.121     &  -26       \\
 & 2 & $4f$--$24l$      & \ce{AB3}--\ce{A2B2}  & \ce{AB}  & -0.045     &  +63       \\
 & 2 & $12k_1$--$12k_1$ & \ce{AB3}--\ce{AB3}   & \ce{B2}  & -0.052     &  -14       \\
 & 2 & $12k_1$--$12k_1$ & \ce{AB3}--\ce{AB3}   & \ce{B2}  & +0.051     &  +88       \\
 & 2 & $12k_1$--$6h_2$  & \ce{AB3}--\ce{A2B2}  & \ce{AB}  & -0.163     & +105       \\
 & 2 & $12k_1$--$12k_2$ & \ce{AB3}--\ce{A2B2}  & \ce{AB}  & -0.091     & +101       \\
 & 2 & $12k_1$--$24l$   & \ce{AB3}--\ce{A2B2}  & \ce{B2}  & -0.178     &  +26       \\
 & 2 & $12k_1$--$24l$   & \ce{AB3}--\ce{A2B2}  & \ce{AB}  & -0.108     &  +96       \\
 & 2 & $6h_1$--$6h_1$   & \ce{A2B2}--\ce{A2B2} & \ce{A2}  & -0.234     & +106       \\
 & 2 & $6h_1$--$12k_2$  & \ce{A2B2}--\ce{A2B2} & \ce{AB}  & -0.269     &  +74       \\
 & 2 & $6h_1$--$24l$    & \ce{A2B2}--\ce{A2B2} & \ce{AB}  & -0.297     &  +59       \\
 & 2 & $6h_2$--$6h_2$   & \ce{A2B2}--\ce{A2B2} & \ce{A2}  & -0.411     &  +87       \\
 & 2 & $6h_2$--$24l$    & \ce{A2B2}--\ce{A2B2} & \ce{AB}  & -0.361     &  +73       \\
 & 2 & $12k_2$--$12k_2$ & \ce{A2B2}--\ce{A2B2} & \ce{B2}  & -0.315     &  +31       \\
 & 2 & $12k_2$--$12k_2$ & \ce{A2B2}--\ce{A2B2} & \ce{AB}  & -0.317     &  +29       \\
 & 2 & $12k_2$--$24l$   & \ce{A2B2}--\ce{A2B2} & \ce{A2}  & -0.258     & +100       \\
 & 2 & $12k_2$--$24l$   & \ce{A2B2}--\ce{A2B2} & \ce{AB}  & -0.302     &  +57       \\
 & 2 & $24l$--$24l$     & \ce{A2B2}--\ce{A2B2} & \ce{AB}  & -0.306     &  +65       \\
 & 2 & $24l$--$24l$     & \ce{A2B2}--\ce{A2B2} & \ce{AB}  & -0.316     &  +55       \\
 & 2 & $24l$--$24l$     & \ce{A2B2}--\ce{A2B2} & \ce{A2}  & -0.275     &  +96       \\
\cmidrule(){2-7}
 & 3 & $6h_1$--$6h_2$   & \ce{A2B2}--\ce{A2B2} & \ce{A2}  & -0.418     &   +1       \\
 & 3 & $12k_2$--$12k_2$ & \ce{A2B2}--\ce{A2B2} & \ce{A2}  & -0.346     &   -1       \\
 & 3 & $24l$--$24l$     & \ce{A2B2}--\ce{A2B2} & \ce{A2}  & -0.372     &   -1       \\
\bottomrule
\end{tabular}

\end{table}

\begin{figure}[!tb]
\centering
\includegraphics{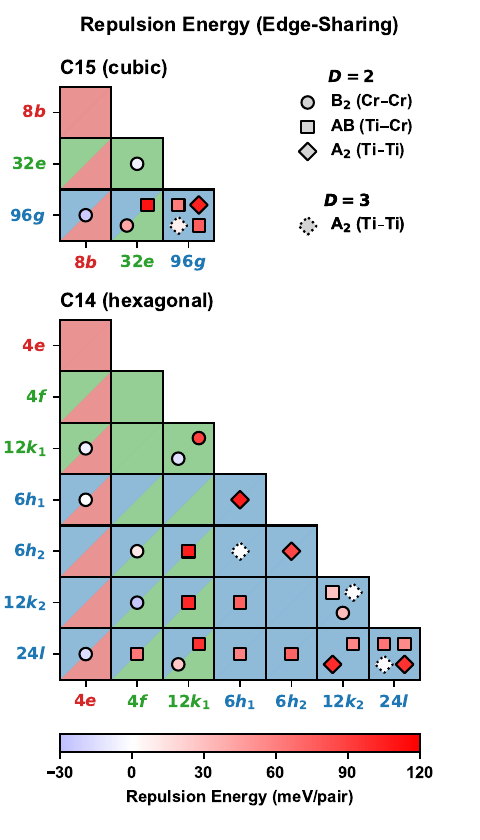}
\caption{Repulsion energies (meV/pair) for all symmetrically distinct edge-sharing hydrogen configurations.
Markers are color-coded based on the repulsion energies, while background colors indicate the types of occupied interstices, where red corresponds to \ce{B4}, green to \ce{AB3}, and blue to \ce{A2B2}.}
\label{fig:repulsion_edges}
\end{figure}

As mentioned in the previous section (Sec.~\ref{sec:results:repulsion_faces}), the dynamically unstable hydrogen atoms in face-sharing configurations migrate into edge-sharing configurations.
Therefore, it is crucial to extend the investigation to edge-sharing configurations to gain deeper insights into hydrogen solubility in the Laves phases.
Unlike the face-sharing configurations, the energetics of edge-sharing configurations were rarely discussed in prior studies.

Table~\ref{tab:repulsion_edges} presents the cumulative binding energies and the repulsion energies of all symmetrically distinct edge-sharing configurations of two hydrogen atoms in the \ce{TiCr2} Laves phases obtained from DFT.
Additionally, Fig.~\ref{fig:repulsion_edges} visually summarizes the results, categorizing them based on the types of the occupied interstices and the shared edges.
All interstice pairs with graph distances of 2 (cf.~\ref{sec:graph_theory}) share a common edge, and a few of those with graph distances of 3 do as well.
The C15 cubic and the C14 hexagonal Laves phases exhibit 8 and
28 symmetrically distinct edge-sharing configurations,
respectively. 

Unlike the face-sharing configurations,
all the edge-sharing configurations are found to be dynamically stable.
Regarding the \ce{A2B2}--\ce{A2B2}
configurations with graph distances of 2, their repulsion energies are substantially lower than those of the face-sharing configurations.
However, they are still substantially high and may therefore be less likely to occur in low hydrogen concentrations.
In contrast, the edge-sharing \ce{A2B2}--\ce{A2B2}
configurations with graph distances of 3 exhibit considerably lower repulsion energies, all below \qty{10}{meV/pair}.
Interestingly, some edge-sharing configurations involving \ce{B4} or \ce{AB3} interstices exhibit negative repulsion energies, indicating attractive interactions, although the hydrogen binding at these individual sites are weaker compared to the \ce{A2B2} interstices (Table~\ref{tab:solution_energy_single}).
Thus, once the \ce{B4} and \ce{AB3} sites start to be occupied at higher hydrogen concentrations, the attractions with these edge-sharing sites may influence hydrogen ordering.

The maximum distance between two hydrogen atoms in the edge-sharing configurations is \qty{2.1}{\angstrom}, which lies at the threshold of the Switendick criterion. Therefore, such configurations may be permissible---confirmed below in the present study.

\subsection{Moment tensor potentials}
\label{sec:results:MTP}

\subsubsection{Error convergence with the MTP complexity level}
\label{sec:results:MTP_level}

\begin{figure}[tb]
    \centering
    \includegraphics{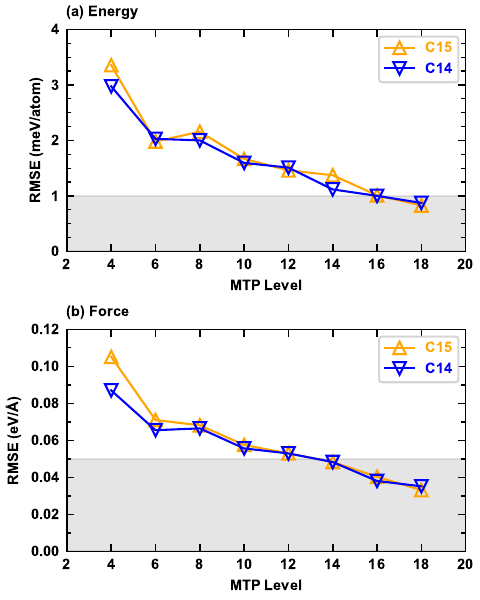}
    \caption{RMSEs in (a) energy and in (b) force as a function of MTP complexity level for the training datasets for dilute hydrogen concentrations.
    The gray-shaded regions show the accuracy we requested in the present study, i.e., 1\,meV/atom for energies and 0.05\,eV/Å for forces.}
    \label{fig:Lev_check}
\end{figure}

The accuracy of any MLIPs is heavily reliant on their hyperparameters. In the fundamental formulation of the MTP, the number of scalar moments increases nearly exponentially with respect to the MTP complexity level~(Sec.~\textcolor{black}{S2} in the SM), so does the flexibility of the potential.
Previous researches indeed demonstrated that the root-mean-square errors (RMSEs) of MTPs from DFT in energy and force decrease with increasing the MTP level~\cite{ou_atomistic_2024,axelrod_learning_2022,jung_high-accuracy_2023,xu_origin_2024}.
On the other hand, increasing the MTP complexity level is directly linked to increases in the  the number of parameters and thus the number of moment operations.
As a result, the computational time required to evaluate the energies, forces, and stresses also grows exponentially with respect to the MTP level and approximately linearly with respect to the number of parameters (Sec.~\textcolor{black}{S2} in the SM).
There is thus a trade-off between predictability and efficiency.
We therefore tested several MTP complexity levels with the training dataset for dilute hydrogen concentrations~(cf.~Sec.~\ref{sec:step-0}).
As shown in Fig.~\ref{fig:Lev_check}, the RMSEs in energy and force decreases as the MTP complexity level, confirming the results in the previous studies.
Specifically, an MTP complexity level of 16 or higher is needed to attain accuracies of 1\,{meV/atom} for energy predictions and 0.05\,{eV/Å} for force predictions.
We thus decided to use the MTPs of level 16 in the present study.

\subsubsection{Performance against training datasets}

\begin{figure*}[!htb]
    \centering
    \includegraphics{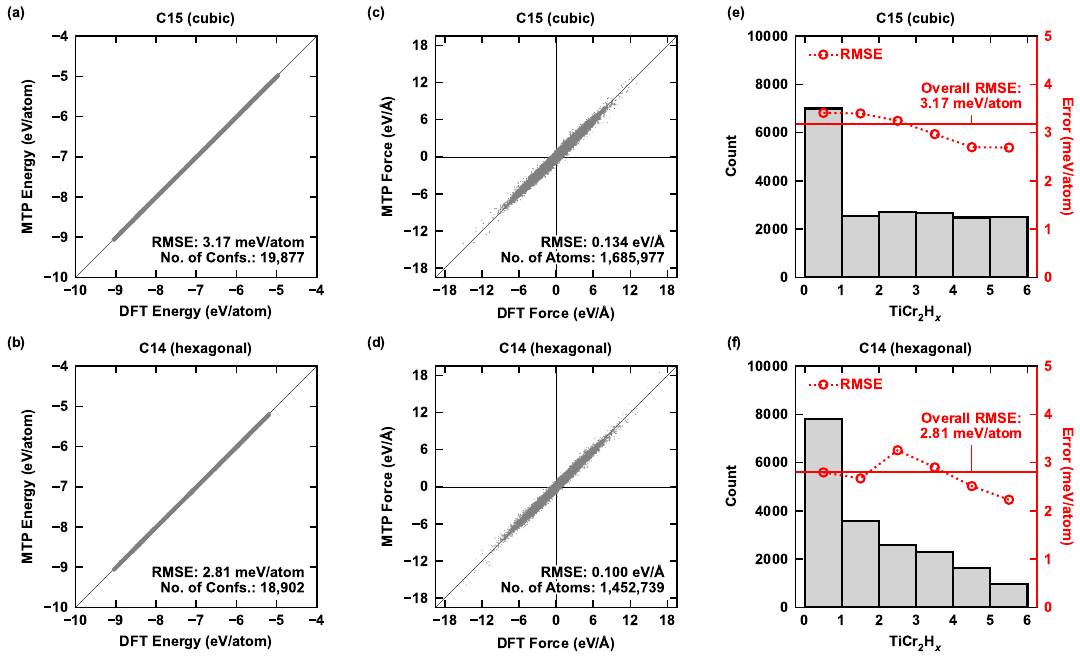}
    \caption{Performance of the trained MTPs in predicting the energies and the forces in different configurations against reference DFT energies and forces for (a, c) C15 and (b, d) C14 configurations, respectively. (e, f) Counts of configurations and the RMSEs for the energies predicted by the MTPs within different hydrogen concentration ranges.}
    \label{fig:Parity}
\end{figure*}

After the active learning schemes described in~Sec.~\ref{subsec:Data-sets}, \num{19877} configurations for the C15 phase (containing \num{1685977} atoms) and \num{18902} configurations for the C14 phases (\num{1452739} atoms) are accumulated.
Fig.~~\ref{fig:Parity}(a,b) show the errors between the MTP-predicted energies from the DFT energies in the training datasets for the C15 and the C14 phases, respectively.
The C15- and C14-MTPs exhibit RMSEs of 3.17\,meV/atom and 2.81\,meV/atom, respectively.
Fig.~\ref{fig:Parity}(c,d) show the errors between the MTP-predicted force components from the DFT values in the training datasets.
The C15- and C14-MTPs achieves RMSEs of 0.134\,eV/Å and 0.100\,eV/Å, respectively, for the individual atomic force component.

The MTPs developed in the present study are designed to simulate the hydrogen solution over a wide range of hydrogen concentrations in the \ce{TiCr2} Laves phases.
Therefore, it is also important to analyze the error statistics in energy predictions across structures with varying hydrogen concentrations.
Fig.~\ref{fig:Parity}(e,f) show the counts of configurations with different hydrogen concentration ranges within the training datasets for the C14 and the C15 phases, respectively.
Notably, there are more configurations at lower hydrogen concentrations ($x < 1$ in \ce{TiCr2H}$_x$) compared to higher concentrations.
This is because low-concentration configurations were systematically and exhaustively generated according to predefined criteria (Sec.~\ref{sec:step-0}), whereas high-concentration configurations were selected using the active-learning scheme.
The number of configurations at $x < 1$ is lower for C15 than for C14 due to the smaller number of symmetrically distinct low-hydrogen configurations for C15 than for C14.
Conversely, at $x > 1$, active learning required a greater number of configurations for C15 than for C14 to ensure comprehensive coverage of the configurational space.
Fig.~\ref{fig:Parity}(e,f) also present the RMSEs
for different hydrogen concentration ranges.
The MTPs achieve similar accuracies across all concentration ranges compared to the overall training datasets, indicating no significant bias toward any particular concentration range.

The MTPs are also validated for test datasets with configurations not included in the training datasets, as detailed in Sec.~\textcolor{black}{S3} in the SM.
The RMSEs for the test datasets are comparable to those for the training datasets, further supporting the robustness of the MTPs.
Sec.~\textcolor{black}{S4.3} in the SM compares the binding energies of a single hydrogen atoms obtained using the MTPs with those obtained using DFT (Sec.~\ref{sec:results:binding_single_H}).
The errors are approximately within \qty{3}{meV/atom}, further demonstrating the accuracy of the MTPs.

\subsection{Enthalpy of formation and occupied interstices}
\label{sec:results:enthalpy_of_formation}

The current MTPs (Sec.~\ref{sec:results:MTP}) are employed for analyzing the enthalpies of formation of \ce{TiCr2H_x} efficiently over a wide range of hydrogen concentrations ($0 < x \le 6$).
Using the MTPs, we identify the interstitial sublattices occupied by hydrogen in the minimum-energy configurations at a given concentration.
These configurations are obtained based on the BHMC simulations detailed in Sec.~\ref{sec:methods:BHMC}.
To ensure the accuracy of the MTPs for higher hydrogen concentrations, additional active-learning is also incorporated during the BHMC procedure.

\subsubsection{C15 cubic TiCr\textsubscript{2}H\textsubscript{\textit{x}}}
\label{sec:results:C15-0-6}

\begin{figure*}[!tb]
    \centering
    \includegraphics[width=\linewidth]{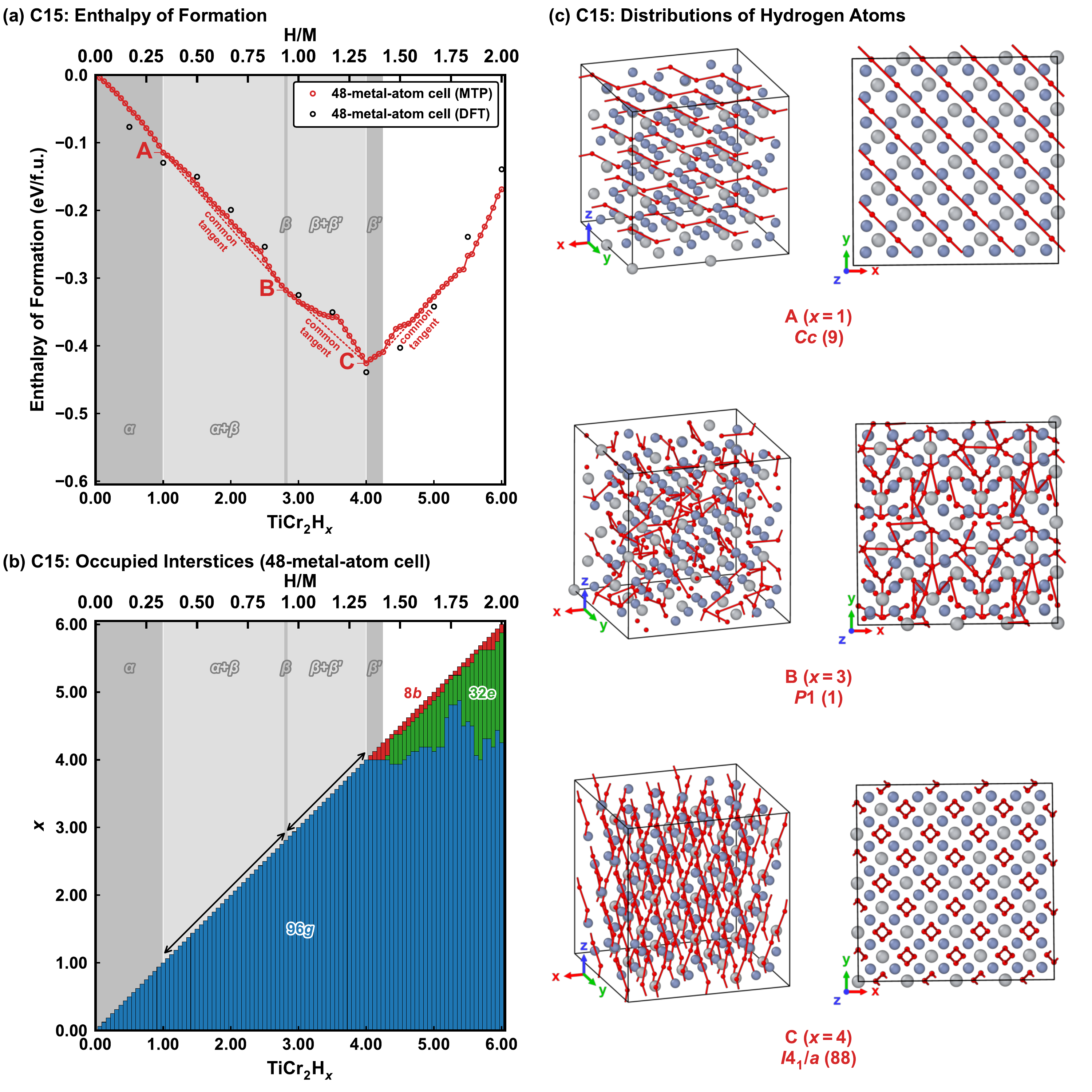}
    \caption{(a) Enthalpy of formation of the C15 cubic \ce{TiCr2H_x} Laves phase for the minimum-energy configurations as a function of $x$.
    Dark- and light-gray-shaded regions indicate single-phase and two-phase states, respectively, with the labels consistent with Johnson and Reilly~\cite{johnson_reaction_1978}.
    (b) Occupied interstices in the minimum-energy configurations. Arrows indicate the concentrations in the two-phase states.
    (c) Distributions of hydrogen atoms (red spheres) in the configurations labeled in (a) as A, B, and C.
The space group types of the given configurations are also shown.
The hydrogen atoms are connected with the cutoff distances of 2.6, 2.6, 2.1\,{\AA} for structures A, B, and C, respectively, to emphasize the hydrogen orderings.}
    \label{fig:C15_sol}
\end{figure*}

Fig.~\ref{fig:C15_sol}(a) presents the enthalpy of formation of the hydrogenated C15 cubic \ce{TiCr2H_x} Laves phase at \qty{0}{K} as a function of hydrogen concentration $x$ as predicted by the MTP.
Fig.~\ref{fig:C15_sol}(b) shows the occupancy of hydrogen atoms at different interstitial sites in the obtained minimum-energy configurations.
At intervals of 0.5 in $x$,
the minimum-energy configurations are also computed using DFT, with the atomic positions re-relaxed while keeping the cell parameters fixed.
The resulting DFT energies (black circles in Fig.~\ref{fig:C15_sol}(a)) are in good agreement with the MTP predictions (red circles) over the investigated hydrogen-concentration range, underscoring the predictive capacity of the MTP.

For low hydrogen concentrations ($x \le 1$), the enthalpy of formation decreases almost linearly, implying that interactions among hydrogen atoms are negligible.
Indeed, in this concentration range, hydrogen preferably occupies the \ce{A2B2} $96g$ tetrahedral interstices without sharing common faces and edges.
The hydrogenated configuration at $x = 1$, at the first cusp on the formation-enthalpy profile, shows a hydrogen ordering with base-centered monoclinic symmetry belonging to space group $Cc$ (9).
As illustrated in Fig.~\ref{fig:C15_sol}(c), this $Cc$ structure features hydrogen atoms ordering along the $\langle 110 \rangle$ direction.
The occupied interstices are connected by vertices but without face- and edge-sharing.
Note that the $Cc$ structure found here is distinct from those showing the same space group or its centrosymmetric supergroup $C2/c$ (15) found previously in experiments by Irodova and Suard~\cite{irodova_evolution_1999} and Kohlmann and Yvon~\cite{kohlmann_revision_2000}, which partly feature edge-sharing among occupied $96g$ interstices.
To the best of the authors' knowledge, the present $Cc$ structure has not been reported in previous experimental and computational studies.
The detailed atomic \textcolor{black}{coordinates} of $Cc$ \ce{TiCr2H} predicted by the MTP and DFT are provided in Table~\ref{tab:hydrides:9}.
\textcolor{black}{The phonon band structure of $Cc$ \ce{TiCr2H} confirms its dynamical stability (Sec.~S5 in the SM).}

\begin{table}[!tbp]
\centering
\scriptsize
\caption{Atomic \textcolor{black}{coordinates} of the $Cc$ \ce{TiCr2H} ordered superstructure relaxed in C15-MTP and DFT with the lattice parameters in C15-MTP of $a = \qty{6.9432}{\angstrom}$, $b = \qty{6.9393}{\angstrom}$, $c =\qty{4.9134}{\angstrom}$, and $\beta = \qty{134.6948}{\degree}$.}
\label{tab:hydrides:9}
\begin{tabular}{cc*6{c}}
\toprule
&
\multirow{2}{*}{Wyckoff} &
\multicolumn{3}{c}{C15-MTP} &
\multicolumn{3}{c}{DFT} \\
& & $x$ & $y$ & $z$ & $x$ & $y$ & $z$ \\
\midrule
Ti & $4a$  &  0.635 &  0.368  &  0.015 &  0.636 &  0.368  &  0.016 \\
Cr & $4a$  &  0.877 &  0.257  &  0.748 &  0.877 &  0.256  &  0.748 \\
Cr & $4a$  &  0.612 &  0.003  &  0.237 &  0.612 &  0.002  &  0.234 \\
 H & $4a$  &  0.948 &  0.058  &  0.018 &  0.948 &  0.057  &  0.020 \\
\bottomrule
\end{tabular}
\end{table}

For $x > 1$, three occupied interstices start sharing one common vertex, forming beyond-one-dimensional vertex-connection-networks.
This causes substantial repulsion among hydrogen atoms and makes the first cusp on the formation-enthalpy profile at $x = 1$.
For $x > 1$, the enthalpy of formation decreases still nearly linearly up to $x \approx 2.5$ but with a smaller slope in magnitude than that in $x \leq 1$.
Above $x \approx 2.5$, a significant drop is found on the formation-enthalpy profile.
The hydrogen concentration at $x \approx 2.5$ marks an onset of the profile,
beyond which a sharp increase of vertex-sharing occupied interstices per hydrogen atom are found~(\ref{sec:long_chain}).

The formation-enthalpy profile in Fig.~\ref{fig:C15_sol}(a) also suggests a two-phase state between $x = 1$ and $x = 2.8125$ (45/16), indicated by the common tangent between the two concentrations.
Thus, the first cusp at $x = 1$ indicates the maximum hydrogen solubility in the C15 cubic Laves phase ($\alpha$), while at higher concentrations it becomes a mixture with the higher-concentration hydride ($\beta$).
In the experimental phase diagram of C15 \ce{TiCr_{1.8}H_{x}} reported by Johnson and Reilly~\cite{johnson_reaction_1978}, the maximum hydrogen solubility in the $\alpha$ phase is $x \approx 1$ at \qty{-100}{\celsius} with the miscibility gap corresponding to the $\alpha$--$\beta$ two-phase state spanning up to $x \approx 2.5$.
The miscibility-gap predicted in the present simulations ($1 < x < 2.8125$) mostly agrees with the experimental data.

On increasing the hydrogen concentration beyond $x = 2.8125$, another sharp drop on the formation-enthalpy profile is observed at $x = 3.625$, leading to the phase transition from $\beta$ to another hydrogenated phase labeled $\beta'$.
Beyond $x = 3.625$, the enthalpy of formation decreases linearly and becomes the lowest at $x = 4$. 
This configuration at this concentration is fully ordered with body-centered tetragonal symmetry belonging to space group $ I4_1/a $ (88).
This $I4_1/a$ phase has a tetragonal unit cell ideally characterized by lattice parameters $a \approx a_{\mathrm{C15}}/\sqrt{2}$ and $c \approx a_{\mathrm{C15}}$, where $a_\mathrm{C15}$ is the lattice parameter of the corresponding C15 phase.
Actually, for the \ce{TiCr2H4} superstructure, MTP predicts $c/a = A \sqrt{2}$ with an anisotropy parameter $A = 1.06$.
As illustrated in Fig.~\ref{fig:C15_sol}(c), in this $I4_1/a$ hydride, hydrogen atoms are arranged along the $c$ direction, i.e., the $\langle 001 \rangle$ direction in the original C15 phase, where nearest two hydrogen atoms in each chain are separated by a distance of \qty{2.1}{\angstrom}.
A closer examination reveals that these hydrogen chains follow a fourfold screw axis ($4_1$) along the $c$ axis.
Notably, hydrogen atoms in this superstructure occupy the $96g$ interstices without sharing common faces, adhering to the Shoemaker--Shoemaker exclusion rule~\cite{shoemaker_concerning_1979}.
This ordered $I4_1/a$ hydrogenated superstructure was also reported in several experimental studies for \ce{HfV2}~\cite{irodova_hydrogen_1981} and \ce{ZrV2}~\cite{didisheim_order-disorder_1981} Laves-phase alloys.
A detailed group--subgroup analysis by Kohlmann~\cite{kohlmann_hydrogen_2020} shows that there is a subgroup series from the original $Fm\bar{3}m$ (227) to first $I4_1/amd$ (141) and finally to $I4_1/a$ (88), and the occupied interstices in the $I4_1/a$ phase is actually the ones split from the $96g$ sites in the original C15 structure.
To the best of the authors' knowledge,  the $I4_1/a$ \ce{AB2H4} hydride has not been reported for \ce{TiCr2}.
It is also worth noting that, although this $I4_1/a$ configuration
is not part of the MTP training dataset, the MTP predicts its energy still in high accuracy with an error of \qty{0.20}{meV/atom} from DFT.
The detailed atomic \textcolor{black}{coordinates} of $I4_1/a$ \ce{TiCr2H4}
predicted by the MTP and DFT are provided in Table~\ref{tab:hydrides:88}.
\textcolor{black}{The phonon band structure of $I4_1/a$ \ce{TiCr2H4} confirms its dynamical stability (Sec.~S5 in the SM).}

Similarly to the $\alpha$--$\beta$ two-phase state, there is also a $\beta$--$\beta'$ two-phase state in $x = 2.8125 \le x \le 4$, as indicated by the common tangent on the formation-enthalpy profile.
The experimental phase diagram of C15 \ce{TiCr_{1.8}H_{x}} by
Johnson and Reilly~\cite{johnson_reaction_1978} shows the corresponding $\beta$--$\beta'$ two-phase state within $2.9 < x < 3.5$ at \qty{-100}{\celsius}, which again mostly agrees with the present computational result.

\begin{table}[tb]
\centering
\scriptsize
\caption{Atomic \textcolor{black}{coordinates} of the $I4_1/a$ (with the origin choice 2, i.e., the origin at the inversion center) \ce{TiCr2H4} ordered superstructure relaxed in C15-MTP and DFT with the lattice parameters in C15-MTP of $a = \qty{5.0667}{\angstrom}$ and $c =\qty{7.6418}{\angstrom}$.}
\label{tab:hydrides:88}
\begin{tabular}{cc*6{c}}
\toprule
&
\multirow{2}{*}{Wyckoff} &
\multicolumn{3}{c}{C15-MTP} &
\multicolumn{3}{c}{DFT} \\
& & $x$ & $y$ & $z$ & $x$ & $y$ & $z$ \\
\midrule
Ti & $4a$  & 0 & 1/4 & 1/8 & 0 & 1/2 & 1/8 \\
Cr & $8d$  & 0 & 0   & 1/2 & 0 & 0   & 1/2 \\
 H & $16f$ & $0.825$ & $0.572$ & $0.063$ & $0.826$ & $0.573$ &  $0.063$ \\
\bottomrule
\end{tabular}
\end{table}

Above $x = 4$, i.e., beyond the experimentally unexplored hydrogen concentrations, \ce{AB3} $8b$ and \ce{B4} $32e$ interstices start to be occupied in the minimum-energy configurations.
Thus, these interstices become energetically more favorable than the remaining \ce{A2B2} $96g$ sites when the hydrogen atoms occupy already one third of the total $96g$ sites.
Interestingly, above $x = 4$, a small amount of $8b$ sites is initially preferred to be occupied prior to the $32e$ sites, even though the latter shows a stronger hydrogen binding for an isolated hydrogen atom (Table~\ref{tab:solution_energy_single}).
This is likely because $32e$ interstices share common faces with the already occupied $96g$ interstices and thus cause a strong repulsion among hydrogen atoms as discussed in Sec.~\ref{sec:results:repulsion_faces}.
Conversely, there are no common faces between $8b$ and $96g$ sites, and hydrogen--hydrogen interactions between edge-sharing $8b$ and $96g$ are found to be attractive (Sec.~\ref{sec:results:repulsion_edges}),
which is why $8b$ sites are occupied first above $x = 4$.
This also highlights the importance to investigate hydrogen configurations at finite concentrations explicitly.

\subsubsection{C14 hexagonal TiCr\textsubscript{2}H\textsubscript{\textit{x}} (low-concentration range)}
\label{sec:results:C14-0-1}

\begin{figure}[tb]
    \centering
    \includegraphics[scale=0.5]{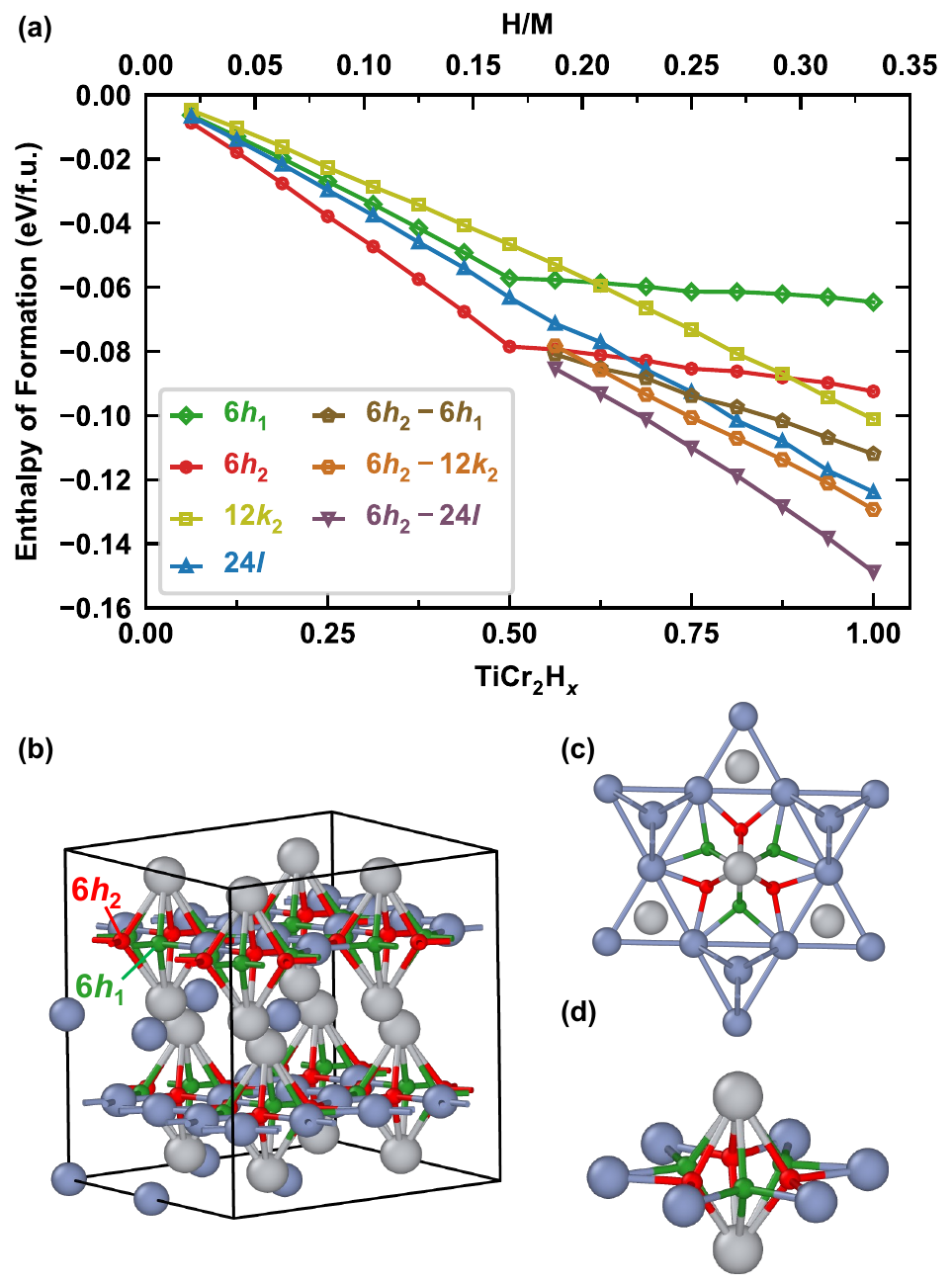}
    \caption{(a) Enthalpy of formation of C14 \ce{TiCr2H_x} occupying different \ce{A2B2} interstices in $0 < x \le 1$.
    (b) $6h_1$ (green) and $6h_2$ (red) interstitial sites.
    (c, d) Isolated views of the single unit of $6h_1$ and $6h_2$ sites sharing a common \ce{Ti}--\ce{Ti} edge.}
    \label{fig:low_C14_solution}
\end{figure}

In this section, we focus on C14 hexagonal \ce{TiCr2H_x} at low hydrogen concentrations ($0 < x \leq 1$).
To gain detailed insights into how hydrogen begins occupying different interstices,
we conducted the BHMC simulations on one or two specific sets of symmetrically distinct interstices.
Fig.~\ref{fig:low_C14_solution}(a) presents the formation-enthalpies of \ce{TiCr2H_x} from these constrained BHMC simulations.

Up to $x = 0.5$, the enthalpy of formation for the $6h_2$-occupied hydrogenated configurations decreases linearly, while above this concentration the slope decreases in magnitude.
Thus, the repulsion among hydrogen atoms becomes substantial above $x = 0.5$.
This trend can be understood well by focusing on the topology of the $6h_2$ sites.
Fig.~\ref{fig:low_C14_solution}(b) illustrates the $6h_2$ sites in \ce{TiCr2}.
There is a unit involving three $6h_2$ interstices with a common \ce{Ti}--\ce{Ti} edge.
As discussed in Sec.~\ref{sec:results:repulsion_edges}, substantial repulsion exists among hydrogen atoms occupying edge-sharing $6h_2$--$6h_2$ interstices with the repulsion energy of 87\,meV/pair.
Consequently, at the low-hydrogen concentrations, only one of the three $6h_2$ sites in each unit can be occupied by hydrogen due to the repulsion.
The total amount of the $6h_2$ interstitial sites corresponds to $x = 1.5$, and thus the maximum amount of hydrogen at $6h_2$ can be only up to $x = 0.5$ at the low concentrations.

Considering the other single-sublattice occupations,
the enthalpy of formation of the $6h_1$-occupied configurations decreases linearly up to $x = 0.5$, and above it the slope decreases in magnitude.
This trend is close to $6h_2$, owing to the
similar topology to $6h_2$ as shown in Fig.~\ref{fig:low_C14_solution}(b).
In contrast, the enthalpies of formation of $12k_2$- and $24l$-occupied configurations decreases linearly even above $x = 0.5$,
indicating that edge-sharing occupations can be avoided for these interstitial sublattices even at $x > 0.5$.
Consequently, when considering hypothetical single-sublattice hydrogen occupation, the most favorable sublattice changes from $6h_2$ to $24l$ at $x \approx 0.75$.

At $x > 0.5$, we also consider hydrogen occupation at another \ce{A2B2} sublattice in addition to $6h_2$ to find further-lower-energy configurations.
At $0.5 < x \le 1.0$, the combination of $6h_2$--$24l$ interstitial sites, featuring vertex-sharing between the two, are found to be energetically more favorable than the single-sublattice occupations as well as the other two combinations, i.e., $6h_2$--$12k_2$ and $6h_2$--$6h_1$.
At $x$ = 1, the hydrogen-preferred occupations in descending order are $6h_2$--$24l$, $6h_2$--$12k_2$, $24l$, and $6h_1$--$6h_2$.
While previous computational studies for C14 Laves phases~\cite{nagasako_first-principles_2002,nong_first-principles_2014,mohammadi_high-entropy_2022} focused on $12k_2$ single-sublattice occupations, the present study highlights the importance of the combination of more than one interstitial sublattice when tracking energetically preferable hydrogen occupations.

\subsubsection{C14 hexagonal TiCr\textsubscript{2}H\textsubscript{\textit{x}} (wide-concentration range) }
\label{sec:results:C14-0-6}

\begin{figure*}[!tb]
    \centering
    \includegraphics[width=\linewidth]{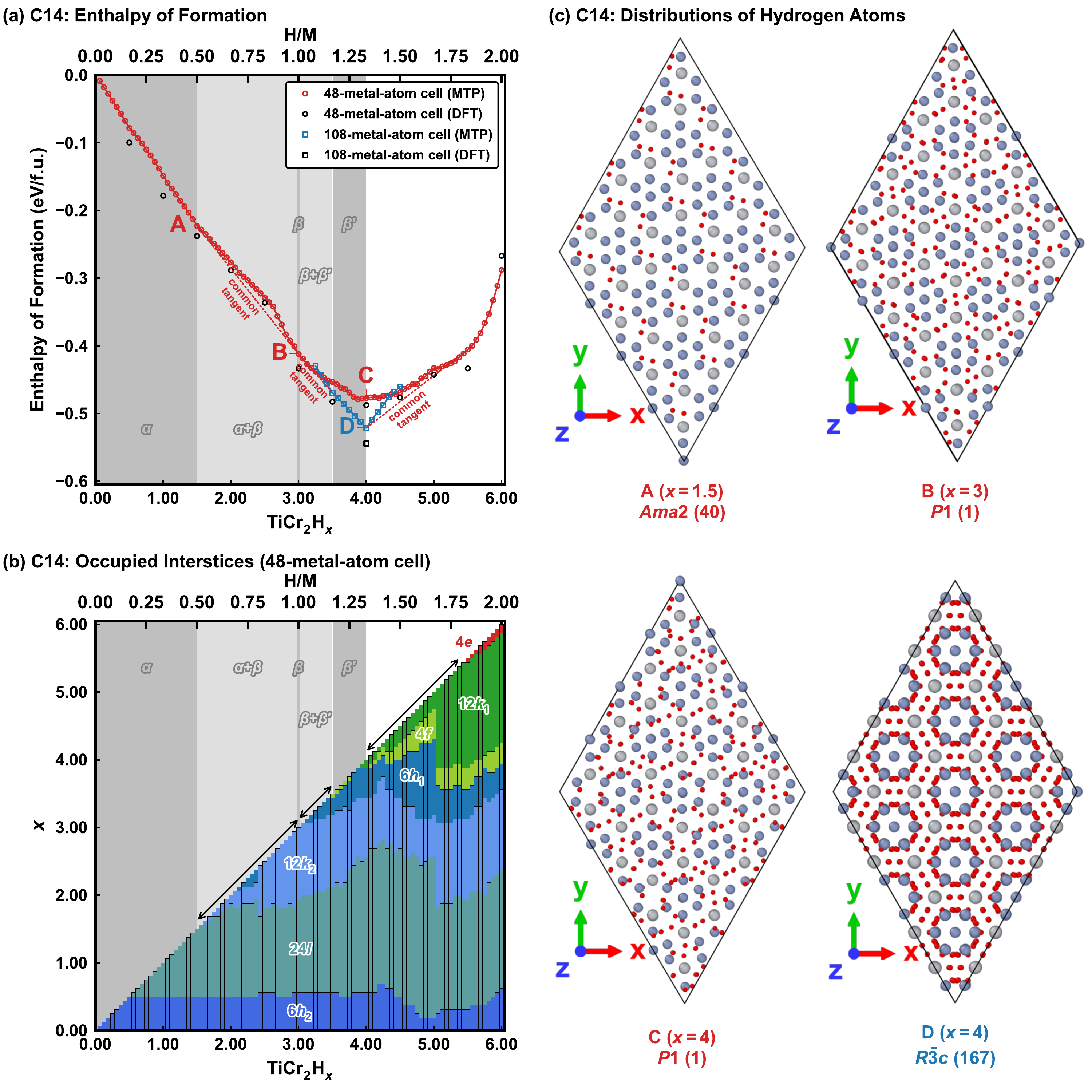}
    \caption{(a) Enthalpy of formation of the C14 hexagonal \ce{TiCr2H_x} Laves phase for the minimum-energy configurations as a function of $x$.
    Dark- and light-gray-shaded regions indicate single-phase and two-phase states, respectively, with the labels consistent with Johnson~\cite{johnson_reaction_1980}.
    (b) Occupied interstices in the minimum-energy configurations.
    Arrows indicate the concentrations in the two-phase states.
    (c) Hydrogen distributions in the configurations labeled in (a) as A, B, C, and D. The space group types of the given configurations are also shown.}
    \label{fig:C14_sol}
\end{figure*}

Fig.~\ref{fig:C14_sol}(a) presents the enthalpy of formation of the hydrogenated C14 hexagonal \ce{TiCr2H_x} Laves phase at \qty{0}{K} as a function of hydrogen concentration $x$ in $0 < x \le 6$ as predicted by the MTP.
Apart from the results with 48-metal-atom cells (circles), the results with larger 108-metal-atom cells (squares) are presented around $x = 4$, as detailed below.
Fig.~\ref{fig:C14_sol}(b) shows the occupancy of hydrogen atoms at different interstitial sites in the obtained minimum-energy configurations with the 48-metal-atom cells.
Like C15 (Sec.~\ref{sec:results:C15-0-6}), the DFT energies (black circles in Fig.~\ref{fig:C14_sol}(a)) agrees well with the MTP predictions (red circles).

At $0 < x \le 0.5$, the enthalpy of formation of C14 \ce{TiCr2H_x} decreases almost linearly with increasing $x$.
In this concentration range, hydrogen occupies the $6h_2$ sites, as discussed already in Sec.~\ref{sec:results:C14-0-1}, and the occupied sites are found to be distant from each other to avoid the repulsion among the hydrogen atoms.
At $x > 0.5$, hydrogen starts to occupy the $24l$ sites.
This change is seen also as a small cusp at $x = 0.5$, where the slope of the formation-enthalpy profile changes very slightly.

The second cusp is found around $x = 1.5$, beyond which the $12k_2$ sites start to be occupied additionally rather than the remaining $24l$ sites.
Thus, in $1.5 \le x \le 3$, the majority of the minimum-energy configurations have hydrogen occupation extended over the $6h_2$--$12k_2$--$24l$ sites.

The light-gray-shaded region in $1.5 \le x \le 3$ in Fig.~\ref{fig:C14_sol}(a) represents the two-phase state consisting of the lower ($\alpha$) and the higher ($\beta$) hydrogen concentration phases, as indicated by the common tangent drawn on the formation-enthalpy profile between the two endpoints.
The highest concentration of $x = 1.5$ in the $\alpha$ phase shows a hydrogen ordering with base-centered orthorhombic symmetry belonging to space group $Ama2$~(40), illustrated in Fig.~\ref{fig:C14_sol}(c).
To the best of the authors' knowledge, the $Ama2$ structure has not been reported in previous experimental and computational studies.
The detailed atomic \textcolor{black}{coordinates} of $Ama2$ \ce{TiCr2H_{1.5}} predicted by the MTP and DFT are provided in Table~\ref{tab:hydrides:C14:Ama2}.
\textcolor{black}{The phonon band structure of $Ama2$ \ce{TiCr2H_{1.5}} confirms its dynamical stability (Sec.~S5 in the SM).}
Within the two-phase range, there is a sharp drop of the formation enthalpy from $x$ = 2.625 (42/16) to $x$ = 2.6875 (43/16), indicating the phase transition,
characterized by a sharp increase of edge-sharing occupied interstices per hydrogen atom (\ref{sec:long_chain}).

In the experimental phase diagram of C14 \ce{TiCr_{1.9}H_x} reported by Johnson \cite{johnson_reaction_1980}, the maximum hydrogen solubility in the $\alpha$ phase is $x \approx 1.6$ at \qty{-100}{\celsius} with the miscibility gap corresponding to the $\alpha$--$\beta$ two-phase state spanning up to $x \approx 2.5$.
The miscibility gap predicted in the present simulations ($1.5 \le x \le 3$) mostly agrees with the experimental data.
This miscibility gap in the C14 phase in the experiments is narrower than the corresponding $\alpha$--$\beta$ miscibility gap in the C15 phase in experiments (Sec.~\ref{sec:results:C15-0-6}).
The present simulations also reproduce this trend in line with the experiments.

\begin{table}[!tbp]
\centering
\scriptsize
\caption{Atomic \textcolor{black}{coordinates} of $Ama2$ \ce{TiCr2H_{1.5}} relaxed in C14-MTP and DFT.
The lattice parameters in C14-MTP are $a = \qty{8.1204}{\angstrom}$, $b = \qty{8.6545}{\angstrom}$, and $c = \qty{4.9521}{\angstrom}$.}
\label{tab:hydrides:C14:Ama2}
\begin{tabular}{cc*6{c}}
\toprule
   & \multirow{2}{*}{Wyckoff} & \multicolumn{3}{c}{C14-MTP} & \multicolumn{3}{c}{DFT} \\
   &       & $x$ & $y$ & $z$ & $x$ & $y$ & $z$ \\
\midrule
Ti & $8c$ &  0.562  &  0.841  &  0.492 & 0.561  &  0.842  &  0.492 \\
Cr & $4a$ &  0      &  0      &  0.013 & 0      &  0      &  0.014 \\
Cr & $4b$ &  1/4    &  0.162  &  0.008 & 1/4    &  0.158  &  0.008 \\
Cr & $4b$ &  1/4    &  0.415  &  0.230 & 1/4    &  0.413  &  0.228 \\
Cr & $4b$ &  1/4    &  0.922  &  0.265 & 1/4    &  0.920  &  0.269 \\
H  & $8c$ &  0.057  &  0.348  &  0.313 & 0.057  &  0.348  &  0.313 \\
H  & $4b$ &  1/4    &  0.228  &  0.678 & 1/4    &  0.226  &  0.677 \\
\bottomrule
\end{tabular}
\end{table}

At $x > 2.6875$ (43/16), the simulations on the 48-metal-atom cell (red circles) suggest the presence of a convex formation-enthalpy profile up to $x = 5$.
At $x > 3$, additional hydrogen atoms occupy mainly the $6h_1$ interstices.
The enthalpy of formation is the lowest at $x \approx 4$.
At $x > 4$, additional hydrogen atoms start to occupy the \ce{AB3} interstices, i.e., $4f$ and $12k_1$ interstices.
At $x = 5$, the formation-enthalpy profile shows again a cusp, corresponding to a sharp increase in $12k_1$ occupancy in Fig.~\ref{fig:C14_sol}(b).
Nevertheless, the hydrogen remains at the given interstices during structural relaxation, indicating their dynamical stability.
At $x \ge 5.5$, hydrogen finally occupies the \ce{B4} $4e$ interstices, and at $x \ge 5.75$, hydrogen occupies face-sharing interstices.

Unlike C15 \ce{TiCr2H_x} (Sec.~\ref{sec:results:C15-0-6}), no sharp peak at $x = 4$ is obtained on the formation-enthalpy profile of C14 \ce{TiCr2H_x} based on the 48-metal-atom cell, and the obtained configuration at this concentration does not show a hydrogen ordering (C in Fig.~\ref{fig:C14_sol}(c)).
In contrast, experiments for C14 \ce{ZrCr2D_{3.8}}~\cite{irodova_orderdisorder_2000,kohlmann_revision_2000} reported a hydrogen ordering at 100\,K.
While Irodova and Suard~\cite{irodova_orderdisorder_2000} originally identified the space group type of this superstructure as $R3$~(146), soon Kohlmann and Yvon~\cite{kohlmann_revision_2000} revised it as $R\bar{3}c$~(167) with inversion symmetry.
The conventional unit cell of the $R\bar{3}c$ superstructure requires a $3 \times 3 \times 3$ expansion of the conventional unit cell of the C14 Laves phase, including 108 metal atoms, which is not commensurate with the $2 \times 2 \times 1$ expansion of the 48-metal-atom cell.
We computed the energy of \ce{TiCr2H4} in this $R\bar{3}c$ superstructure with an idealized hydrogen-occupied interstices using the MTP, allowing relaxation of both atomic positions and cell parameters.
The obtained structure is shown as D in Fig.~\ref{fig:C14_sol}(c).
The enthalpy of formation of this $R\bar3c$ \ce{TiCr2H_x} obtained in MTP (blue square at $x = 4$ in Fig.~\ref{fig:C14_sol}(a)) is substantially more negative than the one obtained from the 48-metal-atom cell by 44\,meV/f.u.
The calculation is also performed in DFT with relaxing the atomic positions again while keeping the cell parameters.
The energy in DFT is only slightly more negative than the MTP energy by \qty{12}{meV}/f.u.~and the enthalpy of formation of the $R\bar3c$ structure in DFT remains substantially more negative than the DFT energy of the configuration of the 48-metal-atom cell by \qty{57}{meV}/f.u.
The detailed atomic \textcolor{black}{coordinates} of $R\bar{3}c$ \ce{TiCr2H4} predicted by the MTP and DFT are provided in Table~\ref{tab:hydrides_Kohlmann}.
One third of each type of \ce{A2B2} interstices, i.e., $6h_1$, $6h_2$, $12_2$, and $24l$, of the original C14 structure is occupied by hydrogen in $R\bar3c$ \ce{TiCr2H4}.
\textcolor{black}{The phonon band structure of $R\bar{3}c$ \ce{TiCr2H4} confirms its dynamical stability (Sec.~S5 in the SM).}

\begin{table}
\centering
\scriptsize
\caption{Atomic \textcolor{black}{coordinates} of $R\bar{3}c$ \ce{TiCr2H4} relaxed in C14-MTP and DFT.
The lattice parameters in C14-MTP are $a = \qty{8.9666}{\angstrom}$ and $c = \qty{25.4204}{\angstrom}$.
The crystal structure is based on \ce{ZrCr2H_{3.8}} at $T = \qty{100}{\kelvin}$ in Kohlmann and Yvon~\cite{kohlmann_revision_2000} with fully filling the hydrogen sites with a fractional occupancy larger than 0.5 while leaving the other hydrogen sites unoccupied.}
\label{tab:hydrides_Kohlmann}
\begin{tabular}{cc*6{c}}
\toprule
   & \multirow{2}{*}{Wyckoff} & \multicolumn{3}{c}{C14-MTP} & \multicolumn{3}{c}{DFT} \\
   &       & $x$ & $y$ & $z$ & $x$ & $y$ & $z$ \\
\midrule
Ti & $36f$ &  0.003 & 0.334 & 0.188 & 0.003 & 0.334 & 0.188 \\
Cr & $36f$ &  0.159 & 0.000 & 0.086 & 0.157 & 0.000 & 0.087 \\
Cr & $18e$ &  0.830 & 0     & 1/4   & 0.829 & 0     & 1/4   \\
Cr & $12c$ &  0     & 0     & 0.173 & 0     & 0     & 0.175 \\
Cr & $6b$  &  0     & 0     & 0     & 0     & 0     & 0     \\
H  & $18e$ &  0.451 & 0     & 1/4   & 0.452 & 0     & 1/4   \\
H  & $36f$ &  0.214 & 0.083 & 0.024 & 0.215 & 0.083 & 0.024 \\
H  & $36f$ &  0.334 & 0.125 & 0.128 & 0.333 & 0.124 & 0.128 \\
H  & $36f$ &  0.092 & 0.210 & 0.151 & 0.091 & 0.209 & 0.151 \\
H  & $18e$ &  0.211 & 0     & 1/4   & 0.213 & 0     & 1/4   \\
\bottomrule
\end{tabular}
\end{table}

Having thus verified the importance of the $R\bar3c$ superstructure, additional BHMC simulations are conducted based on this superstructure with addition or removal of hydrogen atoms in order to extend the formation-enthalpy profile of the $R\bar3c$ superstructure to off-stoichiometric hydrogen concentrations (blue squares in Fig.~\ref{fig:C14_sol}(a)).
The thus obtained enthalpy of formation are found to be more negative than those from the 48-metal-atom cell in $3.5 < x < 4.5$.
Further, a two-phase state is found in $3 < x < 3.5$, where the first end point is on the formation-enthalpy profile at $x = 3$ ($\beta$), while the other at the $R\bar3c$ profile at $x = 3.5$ ($\beta'$).

The experimental phase diagram of C14 \ce{TiCr_{1.9}H_x} by Johnson~\cite{johnson_reaction_1978} shows the corresponding $\beta$--$\beta'$ two-phase state within $3 < x < 3.6$ at \qty{-100}{\celsius}, which again mostly agree with the present computational result.

The obtained $\beta$--$\beta'$ miscibility gap in the C14 phase is narrower than the corresponding gap in the C15 phase (Sec.~\ref{sec:results:C15-0-6}),
also in line with the experiments.

While the 48-metal-atom cell cannot represent the $R\bar3c$ superstructure due to the supercell-shape constraint, the hydrogen occupancy at \ce{TiCr2H4} in this simulation cell may still be insightful.
Table \ref{Tab:Occu_hyd} presents the fractional occupancies of the interstices by hydrogen for C14 \ce{TiCr2H4} obtained from the simulations based on the 48-metal-atom cell.
The $6h_2$ sites show the highest fractional occupancies, followed by the $24l$ sites.
Such a trend is partly consistent with hydrogen-disordered \ce{ZrCr2D_{3.8}} at \qty[unit-color=red]{300}{\kelvin} in neutron-diffraction experiments reported by Irodova and Suard~\cite{irodova_orderdisorder_2000},
despite the difference of constituent elements as well as temperatures.
This suggests that, while the $R\bar3c$ \ce{TiCr2H4} may be the minimum-energy configuration, the critical order-disorder transition temperature of this structure may be below \qty{300}{\celsius}, and the configuration in the 48-metal-atom cell may mimic the disordered configuration at the elevated temperatures.

\begin{table}[tb]
\centering
\scriptsize
\caption{Fractional occupancies of hydrogen at the interstices in the minimum-energy configurations of C14 in the 48-metal-atom cell at $x = 4$ predicted with the MTP.
The values of hydrogen-disordered \ce{ZrCr2D_{3.8}} in neutron-diffraction experiments at 300\,K reported by Irodova and Suard~\cite{irodova_orderdisorder_2000} are shown for comparison.}
\label{Tab:Occu_hyd}
\begin{tabular}{cccc}
\toprule
& Wyckoff & \makecell{\ce{TiCr2H4}\\(present)} & \makecell{\ce{ZrCr2D_{3.8}}\\(Ref.~\cite{irodova_orderdisorder_2000})} \\
\midrule
\ce{B4}
& $4e$    & 0.000 & 0.002 \\
\midrule
\ce{AB3}
& $4f$    & 0.000 & 0.102 \\
& $12k_1$ & 0.020 & 0.044 \\
\midrule
\ce{A2B2}
& $6h_1$  & 0.292 & 0.220 \\
& $6h_2$  & 0.375 & 0.408 \\
& $12k_2$ & 0.292 & 0.276 \\
& $24l$   & 0.333 & 0.304 \\
\bottomrule
\end{tabular}
\end{table}

Apart from the $R\bar3c$ structure, the review of Kohlmann~\cite{kohlmann_hydrogen_2020} referred to other ordered C14-based superstructures for the Laves phases including lanthanide elements~\cite{makarova_interplay_2002}.
For the sake of completeness, we also investigated these hydride structures for \ce{TiCr2} using the MTP and DFT, as detailed in Sec.~\textcolor{black}{S6} in the SM.
They are found to be much more energetically unstable than the minimum-energy configurations obtained in the BHMC simulations, and hence would not appear for \ce{TiCr2H_x}.
Nevertheless, the analysis on these structures provide some insights, e.g., the potential hydrogen occupancy at bipyramidal positions rather than than tetrahedral sites at high hydrogen concentrations.

\section{Conclusions}
\label{sec:Conclusions}

Hydrogen absorption properties in \ce{TiCr2} Laves phases have been investigated at 0\,K using both density functional theory (DFT) and machine-learning interatomic potentials (MLIPs), with the latter being optimized through multiple active-learning schemes.

DFT binding energies reveal that the \ce{A2B2} sites are the most favorable for isolated hydrogen atoms, followed by the \ce{AB3} and the \ce{B4} sites, in both the C14 and C15 Laves phases.
In the C14 hexagonal Laves phase, four symmetrically distinct \ce{A2B2} interstices exhibit substantially different binding energies, influencing the hydrogen occupation in this Laves phase.

The DFT results also reveal that hydrogen atoms at face-sharing interstices experience strong repulsion that can even lead to dynamic instability.
This supports the traditional assumption~\cite{shoemaker_concerning_1979} that hydrogen in the Laves phases does not occupy face-sharing interstices.
Substantial repulsion persists even for hydrogen atoms at edge-sharing interstices, although the magnitude is smaller compared to face-shared sites.
The repulsion among hydrogen atoms may constrain the interstices available for hydrogen atoms in the Laves phases.

To extend and accelerate the analysis toward higher hydrogen concentrations, MLIPs, specifically moment tensor potentials (MTPs)~\cite{novikov_mlip_2020,podryabinkin_active_2017,podryabinkin_mlip-3_2023,shapeev_moment_2016}, have been trained based on DFT.
To achieve accuracy for a wide range of hydrogen concentrations in an effective manner, multiple active-learning schemes have been employed.
The thus trained MTPs predict the energies of the configurations in the training sets with RMSEs of 3.17\,meV/atom and 2.84\,meV/atom for C15 and C14 \ce{TiCr2H_x}, respectively,
for $ 0 \leq x \leq 6 $ ($ 0 \leq \mathrm{H/M} \leq 2$). 

The MTPs, together with basin-hopping Monte Carlo (BHMC) simulations, have allowed us to obtain the formation enthalpies of \ce{TiCr2H_x} in minimum-energy configurations up to a high hydrogen concentration of $x = 6$.
The formation-enthalpy profiles well predict the phase transformations of the hydrogenated \ce{TiCr2H_x} at \qty{0}{K}, reproducing the trends in the experimental phase diagrams at low temperatures~\cite{johnson_reaction_1978,johnson_reaction_1980}.
The hydrogen solubility limits in the low-concentration $\alpha$ phases at \qty{0}{K} are predicted to be $x = 1.0$ and $x = 1.5$ for the C15 and the C14 phases, respectively.
The first and the second hydride phases, i.e., $\beta$ and $\beta'$, at \qty{0}{K} are found around $x = 3$ and $x = 4$, respectively, for both the C15 and the C14 phases.
The good agreement with experiments signifies that the developed MTPs can be used in the future to simulate the pressure--composition--temperature (PCT) diagrams for hydrogen absorption with DFT accuracy.

The BHMC results also elucidate the hydrogen ordering in the minimum-energy configurations.
In the second-hydride $\beta'$ phases at $x = 4$, C15 shows a configuration with the $I4_1/a$ tetragonal symmetry, while C14 shows a configuration with the $R\bar3c$ rhombohedral symmetry.
These configurations were reported in experiments but with different constituent elements, i.e.,  C15 \ce{HfV2}~\cite{irodova_hydrogen_1981}, C15 \ce{ZrV2}~\cite{didisheim_order-disorder_1981}, and C14 \ce{ZrCr2}~\cite{kohlmann_revision_2000}.
In the low-concentration $\alpha$ phases, C15 shows a configuration with the $Cc$ monoclinic symmetry at $x = 1$, while C14 shows a configuration with the $Ama2$ orthorhombic symmetry at $x = 1.5$.
To the best of our knowledge, these configurations had not been reported previously.

Detailed hydrogen occupancies at the different types of interstices in the minimum-energy configurations have also been analyzed.
In both C15 and C14, hydrogen atoms in the minimum-energy configurations occupy mostly the \ce{A2B2} sites up to $x = 4$, above which the \ce{AB3} sites start being occupied.
The C14 phase further reveals substantial preference even among the \ce{A2B2} sites.
Specifically, hydrogen first occupies the $6h_2$ sites up to $x = 0.5$, above which the $24l$ sites start being occupied, consistent with the expectation from the binding energies of the isolated hydrogen atoms.
Further, at low concentrations, hydrogen avoids occupying not only face-sharing but also edge-sharing interstices, supported by the substantial hydrogen repulsion among edge-sharing interstices.

The present findings deepen the comprehensive understanding of the structures and the energetics of hydrogenated \ce{TiCr2} Laves phases.
MLIPs prove to be an efficient method for accelerating the survey, while the developed active-learning schemes maintain their accuracy across a wide range of hydrogen concentrations and also for a wide range of applications.
For example, since training datasets include MD trajectories, the trained MTPs should be robust also for diffusion studies at finite temperatures.
This approach can be directly applied to other Laves-phase alloys, including high-entropy alloys.
Thus, this work provides a roadmap for optimizing hydrogen storage alloys and advancing research on next-generation hydrogen storage materials.

\section*{CRediT authorship contributions}

\textbf{Pranav Kumar}: Conceptualization, Methodology, Software, Validation, Formal Analysis, Investigation, Data Curation, Writing – Original Draft, Visualization.
\textbf{Fritz Körmann}: Writing – Review \& Editing.
\textbf{Blazej Grabowski}: Conceptualization, Resources, Writing – Review \& Editing, Supervision, Project Administration, Funding Acquisition.
\textbf{Yuji Ikeda}: Conceptualization, Software, Visualization, Writing – Review \& Editing, Supervision, Project Administration, Funding Acquisition.

\section*{Declaration of Generative AI and AI-assisted technologies in the writing process}

ChatGPT and Grammarly were used to assist with initial drafting and sentence refinement for certain sections of this manuscript. The authors subsequently reviewed and edited all content as needed and take full responsibility for the final version of the publication.

\section*{Declaration of competing interest}

The authors declare that they have no known competing financial interests or personal relationships that could have appeared to influence the work reported in this paper.

\section*{Acknowledgments}

Pranav Kumar and Yuji Ikeda are funded by the Deutsche Forschungsgemeinschaft (DFG, German Research Foundation), project number 519607530.
Fritz Körmann and Blazej Grabowski acknowledge funding from the European Research Council (ERC) under the European Union’s Horizon 2020 research and innovation programme (grant agreement No.~865855). Fritz Körmann acknowledges support from the Deutsche Forschungsgemeinschaft (DFG, German Research Foundation), project number 541649719.
The authors also acknowledge support by the state of Baden--Württemberg through bwHPC and the DFG through grant no INST 40/575-1 FUGG (JUSTUS 2 cluster) and the SFB1333 (project ID 358283783-CRC 1333/2 2022).

\section*{Data availability}

The developed MTPs along with the corresponding DFT training datasets
are freely accessible on DaRUS.

\section*{Supplementary materials}

Supplementary materials associated with this article can be found online.

\appendix

\section{Local hydrogen configurations}
\label{sec:graph_theory}

\setcounter{table}{0}
\setcounter{figure}{0}

To derive the local hydrogen configurations in Sec.~\ref{sec:step-0}, we utilized space-group symmetry in combination with graph theory.
Specifically, the tetrahedral interstices in the Laves phases are regarded as the nodes of a graph, and the paths to the neighboring face-sharing sites are regarded as the edges.
Each interstice has four face-sharing neighbors, and thus each node has four edges.
Then, we can define the graph distances between any pairs of interstices in the supercell models as the minimum number of edges to be passed to visit from one interstice to the other.
For a given number of hydrogen atoms in the supercell models, we limit the maximum graph distance between the hydrogen atoms.
Specifically, for two-hydrogen configurations, we considered those with a graph distance of up to 4, while for the configurations with more hydrogen atoms, we considered the configurations with a maximum graph distance of up to 3.
Then, by applying symmetry operations, we can enumerate symmetrically inequivalent local hydrogen configurations exhaustively.
The symmetry operations were obtained using the \textsc{SPGLIB} library \cite{togo_spglib_2024}.

Table~\ref{Tab:adj_conf} summarizes the numbers of the thus obtained symmetrically inequivalent configurations.
We considered 1019 and 3766 configurations in total for the C15 cubic and C14 hexagonal Laves phases, respectively.

\begin{table}[!tbp]
\centering
\scriptsize
\caption{Numbers of symmetrically inequivalent local hydrogen configurations in the C15 cubic and the C14 hexagonal Laves phases.
The column ``$N_\mathrm{H}$'' shows the number of hydrogen atoms, the column ``$D$'' shows the maximum graph distances among the hydrogen atoms, and the column ``$N_\mathrm{conf}$'' shows the corresponding numbers of symmetrically inequivalent configurations.} 
\label{Tab:adj_conf}
\begin{tabular}{cccc}
\toprule
& $N_\mathrm{H}$ & $D$ & $N_\mathrm{conf}$ \\
\midrule
C15
& 1 & 0 & 3 \\
& 2 & 1, 2, 3, 4 & 4, 7, 11, 19 \\
& 3 & 2, 3 & 20, 79 \\
& 4 & 2, 3 & 15, 230 \\
& 5 & 2, 3 & 5, 334 \\
& 6 & 3 & 292 \\
\midrule
C14
& 1 & 0 & 7 \\
& 2 & 1, 2, 3, 4 & 11, 24, 39, 69 \\
& 3 & 2, 3 & 65, 302 \\
& 4 & 2, 3 & 45, 874 \\
& 5 & 2, 3 & 13, 1203 \\
& 6 & 3 & 1114 \\
\bottomrule
\end{tabular}
\end{table}

\section{Neighboring hydrogen atoms}
\label{sec:long_chain}

\setcounter{table}{0}
\setcounter{figure}{0}

\begin{figure}[!tb]
    \centering
    \includegraphics{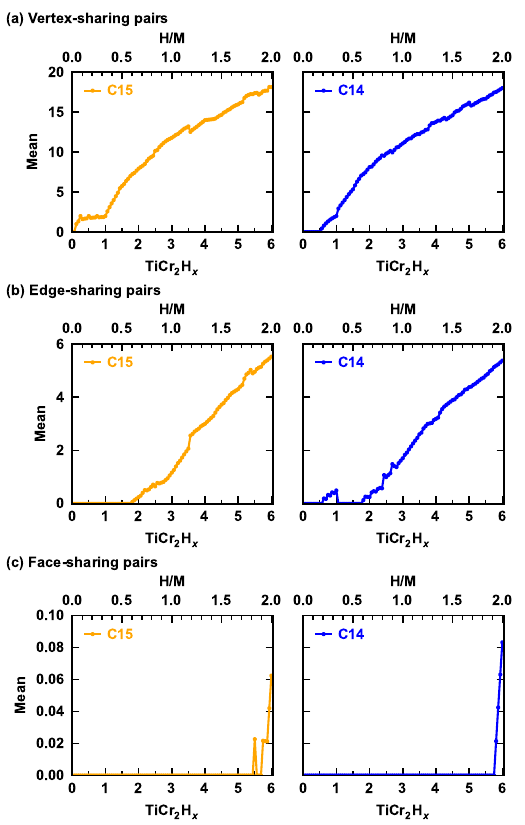}
    \caption{Mean numbers of neighboring hydrogen atoms sharing vertices, edges, and faces in \ce{TrCr2H_x} in the simulations using the 48-metal-atom cells.}
    \label{fig:pairs_avg_2}
\end{figure}

The phase transitions of \ce{TiCr2H_x} discussed in Sec.~\ref{sec:results:enthalpy_of_formation} are associated with qualitative changes in hydrogen-ordering tendencies and may thus be related to local environmental changes around hydrogen atoms.
To capture these changes, the mean numbers of neighboring hydrogen atoms sharing vertices, edges, and faces are analyzed, as shown in Fig.~\ref{fig:pairs_avg_2}.
Indeed, many phase transitions can be characterized by discontinuities in the values and the slopes of the mean numbers of the neighboring hydrogen atoms as a function of hydrogen concentration.
For example, the $\alpha$ and the $\beta$ phases in C14 \ce{TiCr2H_x} are characterized by the number of edge-sharing interstices, which exhibit jumps within $1.5 \le x \le 3$, corresponding to the $\alpha$--$\beta$ two-phase state.

\section{Volume expansion upon hydrogenation}

\setcounter{table}{0}
\setcounter{figure}{0}

\begin{figure}[!htb]
    \centering
    \includegraphics{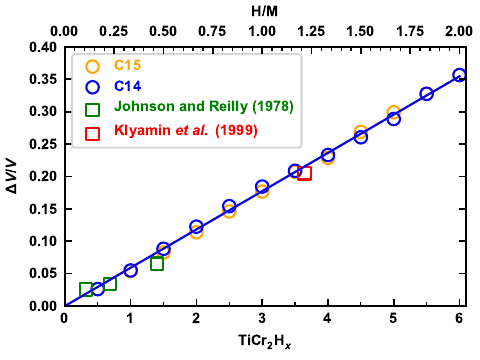}
    \caption{Volume expansion of \ce{TiCr2H_x} as a function of hydrogen concentration predicted by MTPs.
    Computational data are provided for $x$ in increments of 0.5.
    Experimental data are derived from Johnson and Reilly~\cite{johnson_reaction_1978} for the C15 \ce{TiCr_{1.8}H_x} at room temperature and from Klyamkin~\cite{klyamkin_effect_1999} for C14 \ce{TiCr_{1.8}H_{3.40}} at \qty{293}{K} and \qty{200}{MPa}.}
    \label{fig:volume}
\end{figure}

Fig.~\ref{fig:volume} shows the induced volume expansions of \ce{TiCr2H_x} Laves phases as a function of hydrogen concentration obtained using the MTPs.
The data are derived from the minimum-energy configurations identified in the main text.
Both the C15 and C14 phases show similar linear dependence of volume expansion on hydrogen concentration.
The data are also in good agreement with the available experimental values in the literature~\cite{johnson_reaction_1978,klyamkin_effect_1999}.
This agreement further validates the accuracy of the developed MTPs in capturing the volumetric behavior of hydrogenated Laves phases.
The linear fitting shows that the volume increase per hydrogen atom is \qty{2.41}{\angstrom\textsuperscript{3}} for C15 and \qty{2.36}{\angstrom\textsuperscript{3}} for C14 which is consistent with a common value of 2--3\,\AA\textsuperscript{3} in metals~\cite{fukai_metal-hydrogen_2005}.

\bibliographystyle{elsarticle-num-mod}

\end{document}


\begin{frontmatter}
\title{Machine Learning Potentials for Hydrogen Absorption in \texorpdfstring{TiCr\textsubscript{2}}{TiCr2} Laves Phases\\\texorpdfstring{\vspace{\baselineskip}}{}Supplementary Materials}
\end{frontmatter}

\section{Detailed flowcharts for each MTP training scheme}
\label{sec:ext-flow}

\begin{figure}[!htbp]
    \centering
    \includegraphics[scale=0.9]{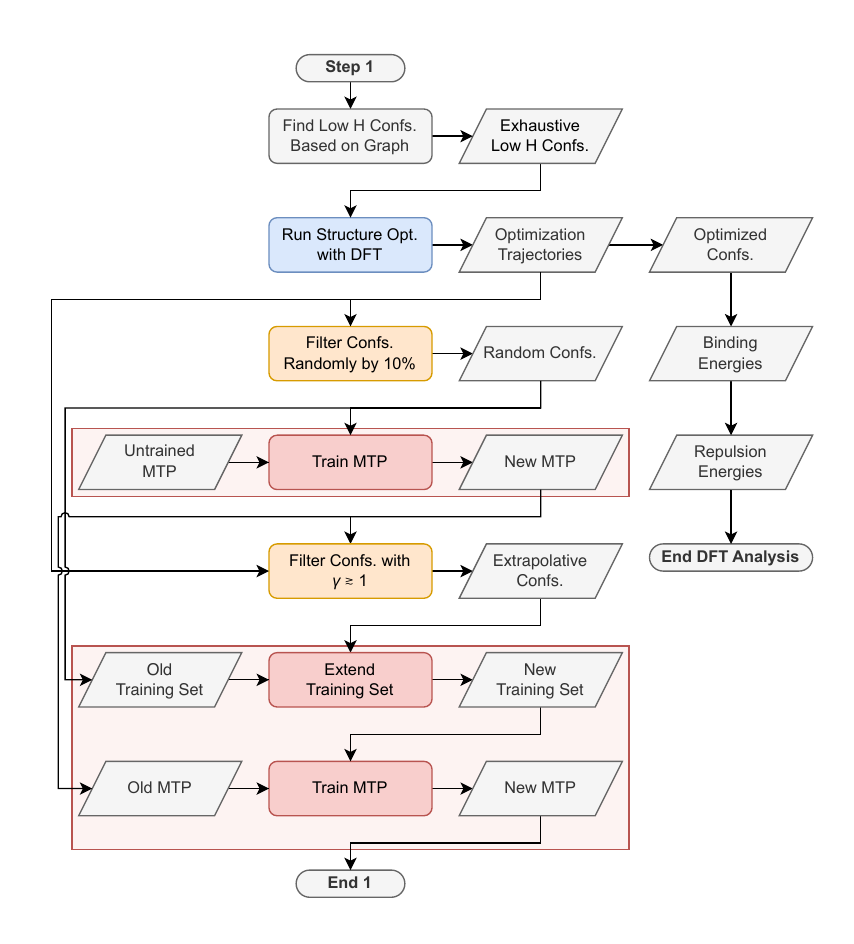}
    \caption{Flowchart depicting the procedure to generate the training dataset and the MTP from low-hydrogen-concentration configurations derived systematically based on graph theory using the using the extrapolation grade $\gamma$. The blue rounded rectangle shows the process involving DFT calculations. The orange rounded rectangle shows the process involving filtering configurations. The red rounded rectangle shows the process training an MTP. The red shaded regions show the processes updating the training dataset and the MTP.}
    \label{fig:step-1}
\end{figure}

\begin{figure}[!htbp]
    \centering
    \includegraphics[scale=0.9]{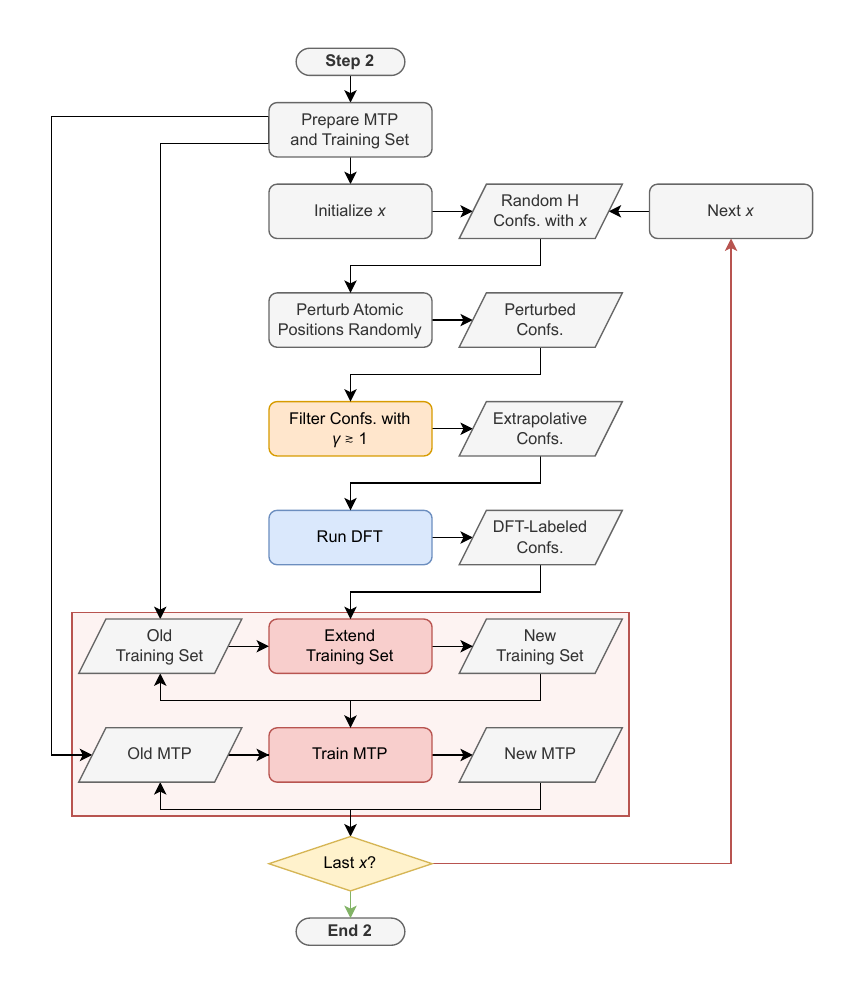}
    \caption{Flowchart depicting the procedure to extend the training dataset and to retrain the MTP with random hydrogen configurations at finite concentrations using the extrapolation grade $\gamma$. The green and the red arrows from the decision symbol means ``yes'' and ``no'', respectively. Other notations are the same as Fig.~\ref{fig:step-1}.}
    \label{fig:step-2}
\end{figure}

\begin{figure}[!htbp]
    \centering
    \includegraphics[scale=0.9]{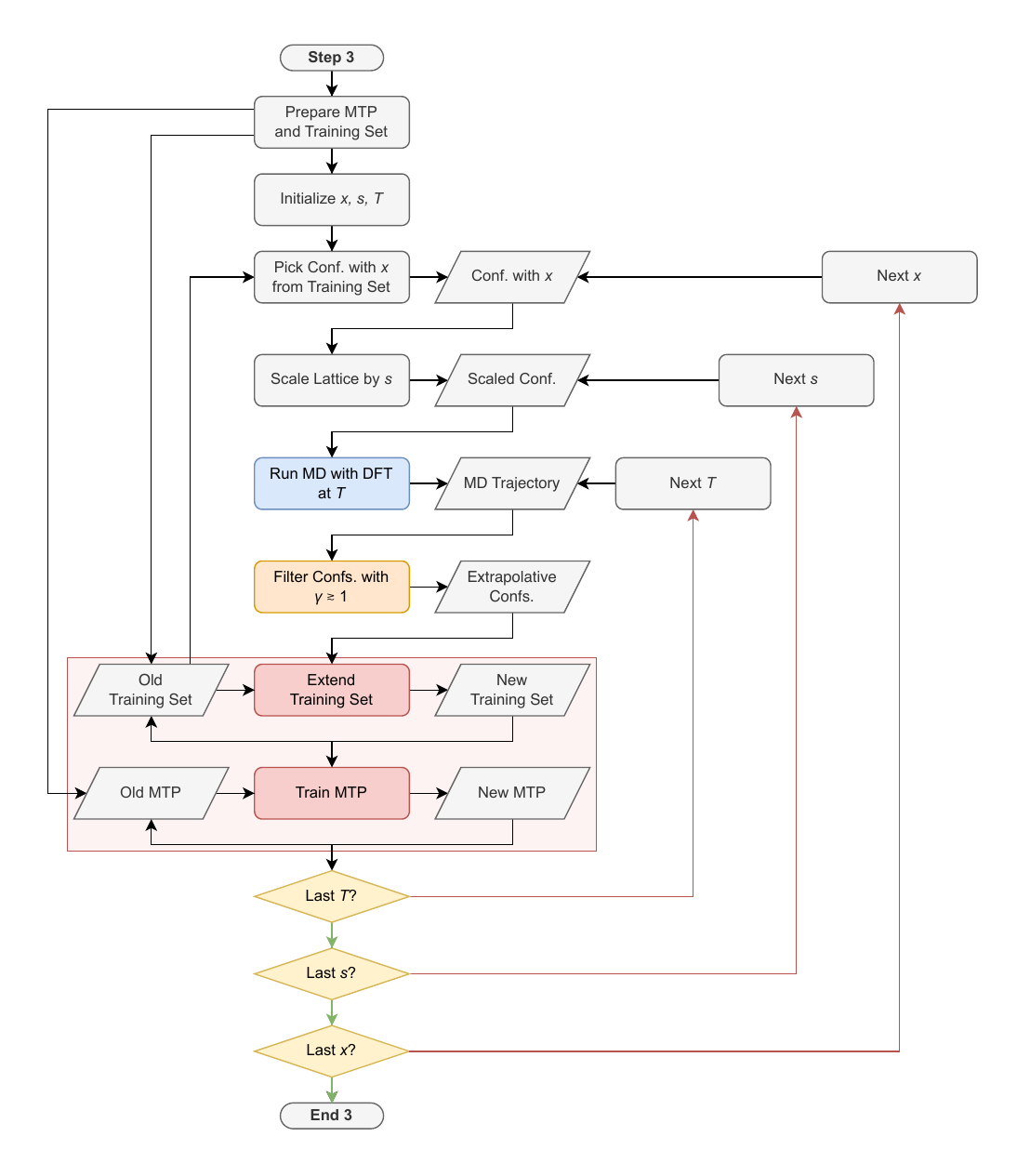}
    \caption{Flowchart depicting the procedure to extend the training dataset and to retrain the MTP based on \textit{ab initio} MD trajectories using the extrapolation grade $\gamma$.
    Notations are the same as Fig.~\ref{fig:step-2}.}
    \label{fig:step-3}
\end{figure}

\begin{figure}[!htbp]
    \centering
    \includegraphics[scale=0.9]{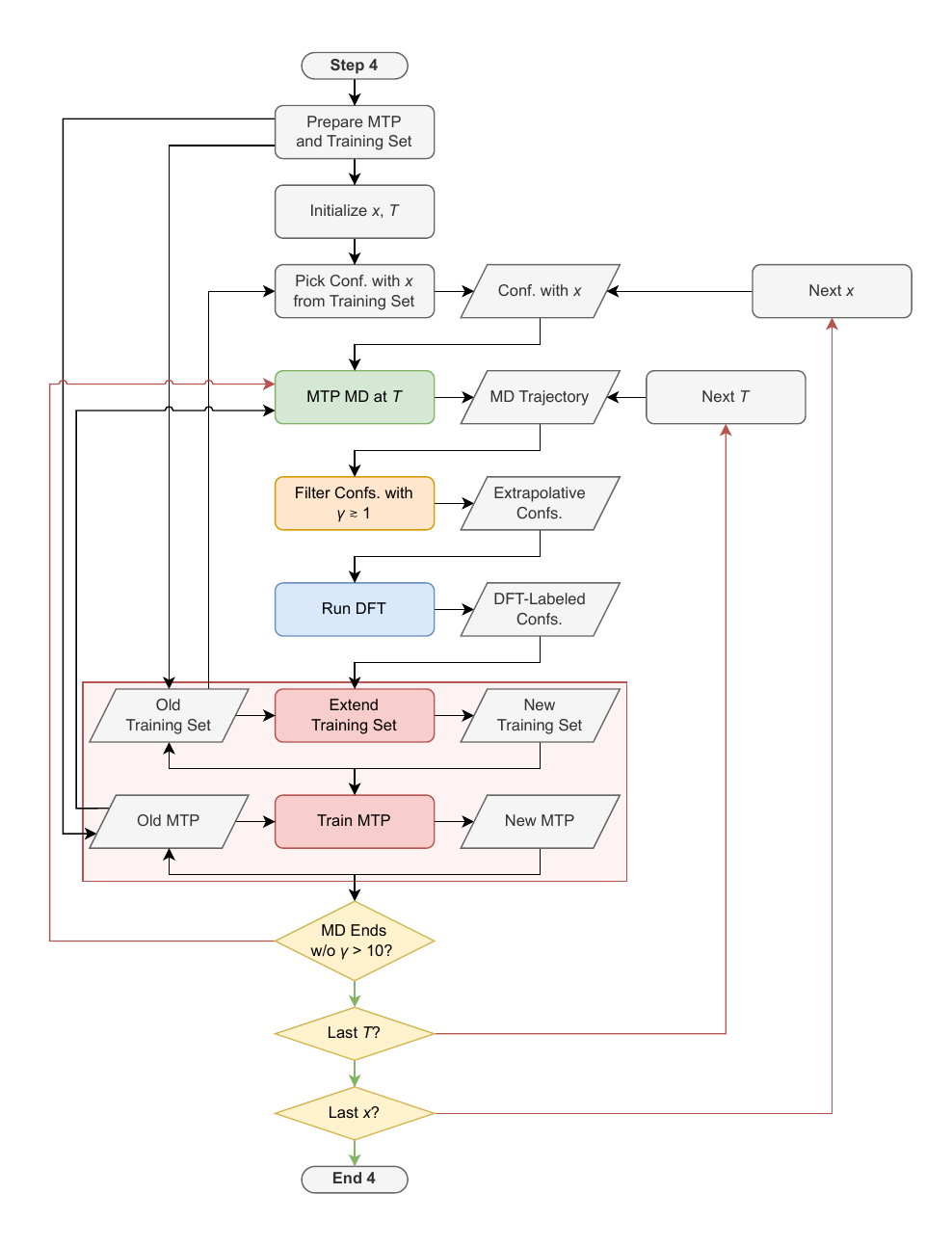}
    \caption{Flowchart depicting the procedure to extend the training dataset and to retrain the MTP based on MD trajectories with the MTP along with active learning. The green rounded rectangle shows the process employing the MTP. Other notations are the same as Fig.~\ref{fig:step-3}.}
    \label{fig:step-4}
\end{figure}

\begin{figure}[!htbp]
    \centering
    \includegraphics[scale=0.9]{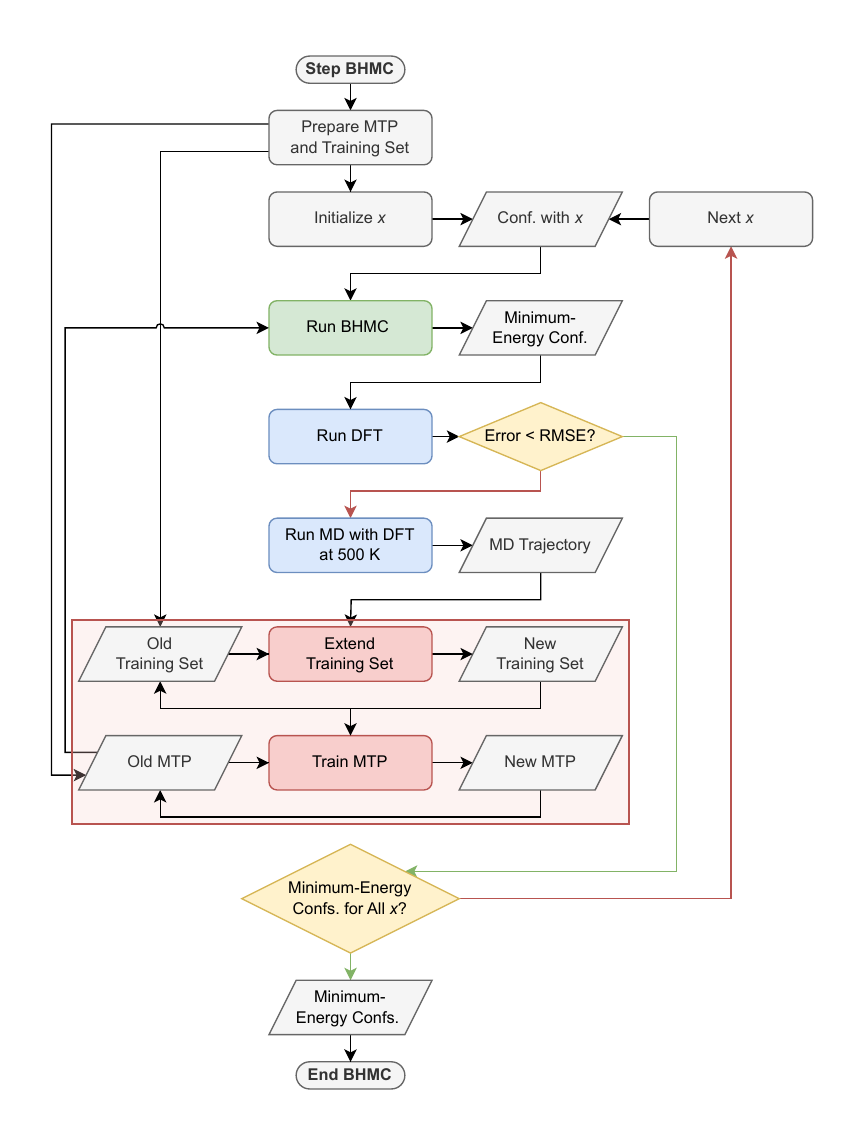}
    \caption{Flowchart depicting the BHMC simulations to find minimum-energy hydrogen configurations for investigated hydrogen concentrations.
    This procedure also involves the retraining procedure of the MTP during the BHMC simulations. Notations are the same as Fig.~\ref{fig:step-4}.}
    \label{fig:step-5}
\end{figure}

Figs.~\ref{fig:step-1}--\ref{fig:step-4} present detailed flowcharts for the active-learning schemes described in Sec.~\textcolor{red}{2.4} in the main text.
Fig.~\ref{fig:step-5} presents the flowchart for the BHMC simulations described in Sec.~\textcolor{red}{2.5} in the main text.


\clearpage
\section{Computational cost of moment tensor potentials}
\label{App:MTP_speed}

\begin{figure}[!htb]
    \centering
    \includegraphics{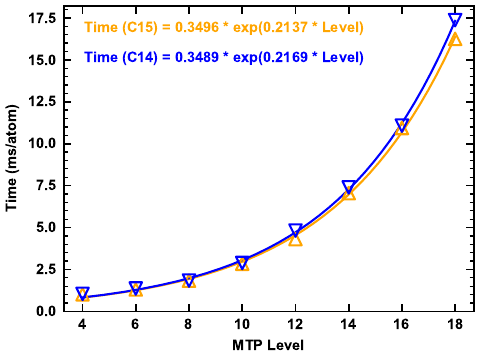}
    \caption{Average evaluation times of C15- and C14-MTPs per atom as a function of MTP level.}
    \label{fig:time}
\end{figure}

As demonstrated in the main text, an MTP with a high complexity level yields lower RMSEs on energies and forces between DFT and MTP predictions.
On the other hand, the evaluation time of an MTP  also increases with its complexity level.
This introduces a trade-off between accuracy and efficiency, making it essential to strike a balance when employing MTPs.
Table~\ref{Tab:time} summarizes the number of parameters in an MTP as a function of MTP complexity level, along with the average evaluation times of C15-MTP and C14-MTP per atom. As visualized in Fig.~\ref{fig:time}, the average evaluation time grows exponentially with increasing MTP complexity.

\begin{table}[!htb]
\centering
\scriptsize
\caption{Number of parameters in MTPs as a function of MTP level.
$Q$ shows the number of Chebyshev polynomials, $\mu$ shows the number of sets of Chebyshev polynomials, and $\zeta$ shows the number of linear coefficients.
Average evaluation times of C15- and C14-MTPs per atom for the corresponding MTP levels are also shown.}
\label{Tab:time}
\begin{tabular}{
ccc*{2}{S[table-format=3.0]}*{2}{S[table-format=2.3,table-auto-round]}}
\toprule
\multirow{2}{*}{Level} &
\multirow{2}{*}{$Q$} &
\multirow{2}{*}{${\mu}$} &
\multicolumn{1}{c}{\multirow{2}{*}{$\zeta$}} &
\multicolumn{1}{c}{\multirow{2}{*}{Parameters}} &
\multicolumn{2}{c}{Time (ms/atom)} \\
&&&&& \multicolumn{1}{c}{C15-MTP} & \multicolumn{1}{c}{C14-MTP} \\
\midrule
 4 &10 &1 &  2 & 96 & 1.023904 & 1.042688 \\
 6 &10 &2 &  5 &189 & 1.309363 & 1.363451 \\
 8 &10 &2 &  9 &193 & 1.832869 & 1.837469 \\
10 &10 &3 & 16 &290 & 2.843426 & 2.88702  \\
12 &10 &3 & 29 &303 & 4.310956 & 4.819368 \\
14 &10 &4 & 52 &416 & 7.049801 & 7.391734 \\
16 &10 &4 & 92 &456 &10.936056 &11.074666 \\
18 &10 &5 &163 &617 &16.246614 &17.365337 \\
\bottomrule
\end{tabular}
\end{table}

\clearpage
\section{Validation of the MTPs on the test datasets}

Our trained MTPs were also validated for configurations not included in the training datasets.
Specifically, 95 hydrogen concentrations within $0 < x \le 6$ with a step of $1/48$ were considered, excluding $x = 1/48$ (one hydrogen atom in the 48-atom cell).
For each concentration, 21 configurations with randomly distributed hydrogen atoms were first generated, resulting in \num{1995} configurations (approximately 10\% of the training datasets).
These configurations were subsequently relaxed with respect to both atomic positions and cell parameters using our trained MTPs.
Next, all \num{1995} relaxed configurations underwent MD simulations within the \textit{NVT} ensemble using our trained MTPs at \qty{750}{K}, with a time step of \qty{1}{fs} and a total simulation time of \qty{0.1}{ns}.
Following the MD simulations, the final configuration of each trajectory was further relaxed with respect to atomic positions and lattice parameters.
Finally, we performed single-point DFT calculations on these relaxed structures, creating test datasets to assess the validity of the trained MTPs.

\begin{figure}[!htbp]
    \centering
    \includegraphics{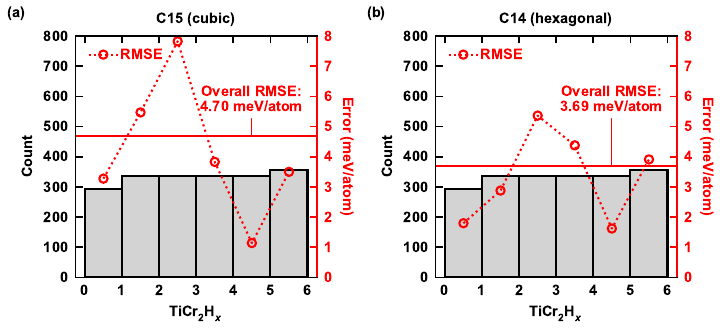}
    \caption{Counts of configurations in the test datasets and the RMSEs for the energies predicted by the MTPs within different hydrogen concentration ranges.}
    \label{fig:performance_test}
\end{figure}

Fig.~\ref{fig:performance_test} shows the RMSE distributions of C15- and C14-MTPs for different concentration ranges in $0 < x \leq 6$ for the above-described test datasets.
The RMSEs for the test datasets are \qty{4.70}{meV/atom} for C15-MTP and \qty{3.69}{meV/atom} for C14-MTP, which are comparable to RMSEs for the training datasets, \qty{3.17}{meV/atom} and \qty{2.81}{meV/atom}, respectively.
This indicates that the trained MTPs are robust for use in hydrogen configurations even not in the training datasets.

\clearpage

\section{Binding energies of a single hydrogen atom in more detail}

\subsection{Impact of optimization schemes}

The hydrogen binding energies obtained in atomistic simulations substantially depend on the structure-optimization scheme.
To highlight its significance, we tested the following three different optimization schemes for the cell parameters and the atomic coordinates:

\begin{itemize}
  \item[(a)] \textbf{Frozen-host:} Both the cell parameters and the coordinates of the metal atoms are fixed to those of the pure \ce{TiCr2} Laves phases, and only the coordinates of the hydrogen atom is allowed to relax.
  \item[(b)] \textbf{Fixed-cell:} The cell parameters are fixed to those of the pristine \ce{TiCr2} Laves phase, but all the atomic coordinates are relaxed after the placing the hydrogen atom.
  \item[(c)] \textbf{Full-relaxation:} Both the cell parameters and the atomic coordinates are relaxed after placing the  hydrogen atom (scheme discussed in the main text).
\end{itemize}

Table \ref{tab:solution_energy_single_vs_opt} presents the hydrogen binding energies obtained with the three schemes in DFT.
As naturally expected, the differences between the fixed-cell and the full-relaxation schemes are relatively small, while the values in the frozen-host scheme are significantly higher than the other two schemes.
This may explain the apparent discrepancy between our values and those in Li~\textit{et al.}~\cite{li_hydrogen_2009} for the C15 \ce{TiCr2} shown in Table~2 in the main text.
Specifically, Li~\textit{et al.}~\cite{li_hydrogen_2009} obtained the binding energy at the \ce{B4} site (\qty{+1.68}{eV/H}) probably with the frozen-host scheme, as stated in their paper, while those at the \ce{AB3} site (\qty{-0.02}{eV/H}) and the \ce{A2B2} site (\qty{-0.22}{eV/H}) were possibly obtained with the full-relaxation scheme.

\begin{table}[htbp]
\centering
\scriptsize
\caption{Binding energies $E_\mathrm{b}$ (eV/H) of a single hydrogen atom at the interstices of the C15 cubic and the C14 hexagonal \ce{TiCr2} Laves phases obtained from the DFT simulations using several optimization schemes for the simulation cell and the atomic coordinates.
Note that the full-relaxation scheme is discussed in the main text.}
\label{tab:solution_energy_single_vs_opt}
\begin{tabular}{
ccc
*3{S[table-format=3.4,retain-explicit-plus]}
}
\toprule
& Type & Site & \multicolumn{1}{c}{Frozen-host}&\multicolumn{1}{c}{Fixed-cell} & \multicolumn{1}{c}{Full-relaxation} \\
\midrule
C15                  & \ce{B4}   & $8b$    & +1.669 & +0.637 & +0.599 \\
\cmidrule{2-6}
                     & \ce{AB3}  & $32e$   & +0.647& +0.001 & -0.016 \\
\cmidrule{2-6}
                     & \ce{A2B2} & $96g$   & -0.006& -0.206 & -0.217 \\
\midrule
C14                  & \ce{B4}   & $4e$    & +1.426 & +0.604 & +0.566 \\
\cmidrule{2-6}
                     & \ce{AB3}  & $4f$    & +0.546 & +0.101 & +0.077 \\
                     &           & $12k_1$ & +0.299& +0.001 & -0.019 \\
\cmidrule{2-6}
                     & \ce{A2B2} & $6h_1$  &-0.012 & -0.155 & -0.170 \\
                     &           & $6h_2$  & -0.071& -0.233 & -0.249 \\
                     &           & $12k_2$ & +0.017& -0.158 & -0.173 \\
                     &           & $24l$   & -0.017& -0.171 & -0.185 \\
\bottomrule
\end{tabular}
\end{table}

\subsection{Impact of zero-point energies}

Table~\ref{tab:solution_energy_single_ZPE} presents the zero-point energies (ZPEs) and their impact on the binding energies of a single hydrogen atom obtained with DFT.
The ZPE calculations were conducted within the harmonic approximation based on the finite-displacement method with a displacement of \qty{0.015}{\angstrom} as implemented in VASP~\cite{furthmuller_dimer_1996,kresse_efficiency_1996,kresse_ultrasoft_1999}.
The ZPEs of a single hydrogen atom in \ce{TiCr2} were computed employing the 48-metal-atom cell with fixing the metal atoms because they are are much heavier than the hydrogen atom and therefore approximately immobile with respect to the hydrogen vibration.
The ZPE of an \ce{H2} molecule was obtained as \qty{0.266}{eV/molecule}, in good agreement with the experimental value by Irikura (\qty{0.270}{eV/molecule})~\cite{irikura_experimental_2007,irikura_erratum_2009}.

In both the C15 cubic and the C14 hexagonal Laves phases, the ZPE of a hydrogen atom is the largest at the \ce{B4} interstitial site, followed by the \ce{AB3} and the \ce{A2B2} sites in descending order.
This is likely because the \ce{B4} site is smaller than the others and therefore exhibits a stronger repulsion from the surrounding metal atoms when the hydrogen atom vibrates, resulting in higher vibrational frequencies.
The obtained ZPEs are in close agreement with those obtained semi-empirically for C15 \ce{TiCr_{1.85}} (\qtylist{0.235;0.215}{eV/H} for the $32e$ and the $96g$ sites, respectively)~\cite{fernandez_empirical_1999}.

The ZPEs per hydrogen atom are substantially larger than the value in an \ce{H2} molecule (\qty{0.133}{eV/H}), and hence the binding energies become more positive after the ZPE correction, and thus the hydrogen solubility becomes lower.
However, the order of the binding energies among the interstitial sites in each Laves phase is not substantially affected by the ZPEs.
Specifically, in both the C15 cubic and the C14 hexagonal Laves phases, the \ce{A2B2} interstices are the most energetically favorable for hydrogen, followed by the \ce{AB3} and the \ce{B4} interstices in descending order, and particularly for the C14 hexagonal phase,
the hydrogen solubility at the seven distinct interstices is ordered as $6h_2 > 24l \approx 12k_2 \approx 6h_1 > 12k_1 > 4f > 4e$.
Thus, the ZPE correction would not affect the qualitative findings in the present study such as the presence of the novel hydride phases for \ce{TiCr2}.

Note that the ZPEs in DFT reported for C15 \ce{TiCr2} by Li \textit{et al.}~\cite{li_hydrogen_2009} (\qtylist{0.13;0.11;0.09}{eV/H} for the $8e$, the $32e$, and the $96g$ sites, respectively) are significantly smaller than our values as well as the semi-empirical ones~\cite{fernandez_empirical_1999}, which inverts the ZPE correction and shifts the binding energies toward more negative values.
We attribute this discrepancy to an analytical error in the previous work.

\begin{table}[htb]
\centering
\scriptsize
\caption{ZPEs and their impact $\Delta_\mathrm{ZPE}$ on the binding energies $E_\mathrm{b}$ (eV/H) of a single hydrogen atom at the interstices of the C15 cubic and the C14 hexagonal \ce{TiCr2} Laves phases obtained from the DFT simulations.}
\label{tab:solution_energy_single_ZPE}
\begin{tabular}{
ccc
*4{S[table-format=3.4,retain-explicit-plus]}
}
\toprule
&
Type &
Site &
\multicolumn{1}{c}{ZPE} &
\multicolumn{1}{c}{$\Delta_\mathrm{ZPE}$} &
\multicolumn{1}{c}{$E_\mathrm{b}$ (w/o~ZPE)} &
\multicolumn{1}{c}{$E_\mathrm{b}$ (w/~ZPE)} \\
\midrule
C15                  & \ce{B4}   & $8b$    &0.319& +0.186 & +0.599 & +0.785 \\
\cmidrule{2-7}
                     & \ce{AB3}  & $32e$   &0.268& +0.135 & -0.016 & +0.119 \\
\cmidrule{2-7}
                     & \ce{A2B2} & $96g$   &0.254& +0.121 & -0.217 & -0.096 \\
\midrule
C14                  & \ce{B4}   & $4e$    &0.312& +0.179 & +0.566 & +0.745 \\
\cmidrule{2-7}
                     & \ce{AB3}  & $4f$    &0.279& +0.146 & +0.077 & +0.223 \\
                     &           & $12k_1$ &0.265& +0.132 & -0.019 & +0.113 \\
\cmidrule{2-7}
                     & \ce{A2B2} & $6h_1$  &0.247& +0.114 & -0.170 & -0.057 \\
                     &           & $6h_2$  &0.255& +0.121 & -0.249 & -0.128 \\
                     &           & $12k_2$ &0.256& +0.122 & -0.173 & -0.050 \\
                     &           & $24l$   &0.253& +0.120 & -0.185 & -0.065 \\
\bottomrule
\end{tabular}
\end{table}

\subsection{Comparison with MTP values}

Table~\ref{tab:DFT_vs_MTP} presents the binding energies of a single hydrogen atom obtained from the MTPs and compares them with the DFT values (Table~\textcolor{red}{2} in the main text).
In each method, the binding energies are obtained in the 48-metal-atom cells by relaxing both atomic positions and cell parameters.

The MTPs reproduce the trends found with DFT well.
Specifically, in both the C15 cubic and the C14 hexagonal Laves phases, the \ce{A2B2} interstices are the most energetically favorable for hydrogen, followed by the \ce{AB3} and the \ce{B4} in descending order, and for the C14 hexagonal phase, the $6h_2$ \ce{A2B2} interstices are the most energetically favorable for hydrogen, followed by the $24l$ interstices.

Note that, although the RMSEs of the MTPs with respect to the training datasets are as small as \qty{3}{meV/atom} (Sec.~\textcolor{red}{3.3.2} in the main text), the errors in the hydrogen binding energies are scaled by the total number of atoms in the supercells.
The same issue also obscures other supercell-dependent properties, such as vacancy formation energies.
This highlights a general challenge for MLIPs in reproducing supercell-dependent properties further accurately.

\begin{table}[htb]
\centering
\scriptsize
\caption{Binding energies $E_\mathrm{b}$ (eV/H) of a single hydrogen atom at the interstices of the C15 cubic and the C14 hexagonal \ce{TiCr2} Laves phases, as obtained from the MTP and the DFT simulations.
Errors in the MTPs are also shown both in eV/cell and in eV/atom.}
\label{tab:DFT_vs_MTP}
\begin{tabular}{ccc*4{S[table-format=3.3,retain-explicit-plus=true]}}
\toprule
&
\multirow{2}{*}{Type} &
\multirow{2}{*}{Site} &
\multicolumn{2}{c}{$E_\mathrm{b}$ (\unit{eV/H})} &
\multicolumn{2}{c}{$E_\mathrm{b}^\mathrm{MTP} - E_\mathrm{b}^\mathrm{DFT}$} \\
& & &
\multicolumn{1}{c}{DFT} &
\multicolumn{1}{c}{MTP} &
\multicolumn{1}{c}{(eV/cell)} &
\multicolumn{1}{c}{(eV/atom)} \\
\midrule
C15
& \ce{B4}    & $8b$       &  0.599 &  0.647 & +0.048 & +0.001 \\
\cmidrule{2-7}
 & \ce{AB3}   & $32e$      & -0.016 &  0.104 & +0.120 & +0.002 \\
\cmidrule{2-7}
 & \ce{A2B2}  & $96g$      & -0.217 & -0.066 & +0.151 & +0.003 \\
\midrule
C14
 & \ce{B4}    & $4e$       &  0.566 &  0.675 & +0.109 & +0.002 \\
\cmidrule{2-7}
 & \ce{AB3}   & $4f$       &  0.077 &  0.169 & +0.092 & +0.002 \\
 &            & $12k_1$    & -0.019 &  0.074 & +0.093 & +0.002 \\
\cmidrule{2-7}
 & \ce{A2B2}  & $6h_1$     & -0.170 & -0.100 & +0.070 & +0.001 \\
 &            & $6h_2$     & -0.249 & -0.138 & +0.111 & +0.002 \\
 &            & $12k_2$    & -0.173 & -0.076 & +0.097 & +0.002 \\
 &            & $24l$      & -0.185 & -0.107 & +0.079 & +0.002 \\
\bottomrule
\end{tabular}
\end{table}

\clearpage

\section{Phonon analysis of newly predicted C15-based and C14-based hydrides}

\begin{figure}[!ht]
    \centering
    \includegraphics{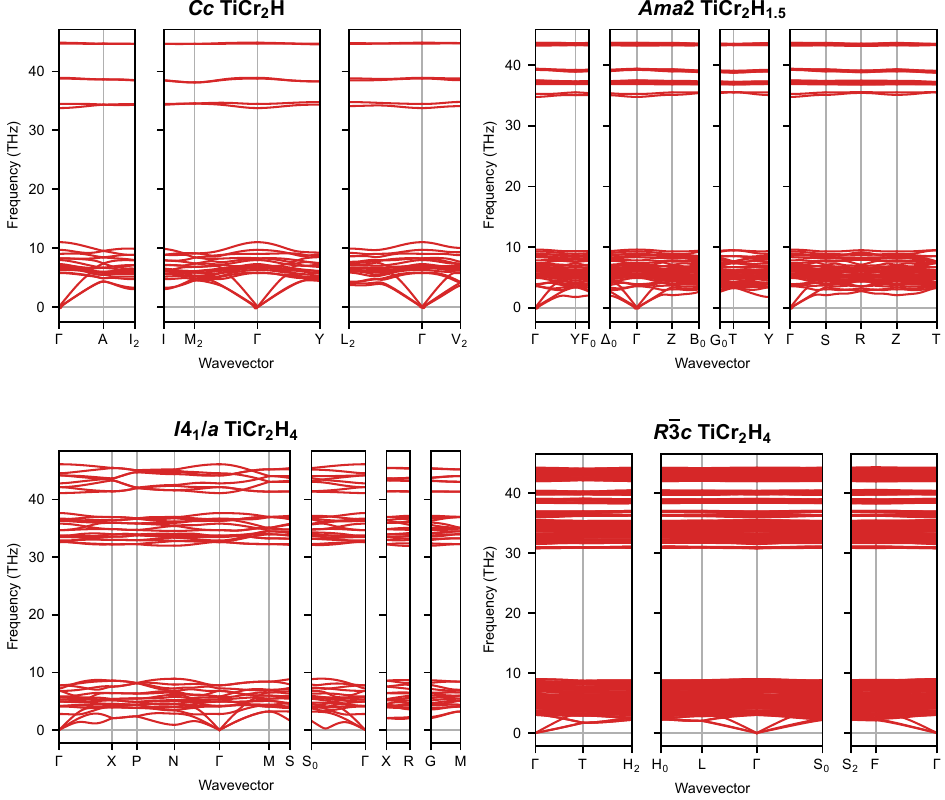}
    \caption{Phonon band structures of the C15 and C14-based hydrides of \ce{TiCr2} calculated with DFT.
    The labels for the high-symmetry wavevectors follow Hinuma~\textit{et al.}~\cite{hinuma_band_2017}}
    \label{fig:Phonon}
\end{figure}

We analyzed the phonon band structures of the newly found C15- and C14-based hydrides of \ce{TiCr2} at \qty{0}{K} within the harmonic approximation in DFT to analyze their dynamical stability.
The finite-displacement method implemented in Phonopy~\cite{togo_implementation_2023} was applied with the amplitude of the displacements set to \qty{0.005}{\text{\AA}}.
For each hydride, a \(2 \times 2 \times 2\) supercell of its primitive unit cell was considered, and the k-point meshes in the reciprocal space were set so that the corresponding k-point-mesh densities become similar to those used for the 48-metal-atom cells.
Fig.~\ref{fig:Phonon} presents the thus obtained phonon band structures.
None of the hydrides shows imaginary modes, indicating their dynamical stability at \qty{0}{K}.

\color{black}

\clearpage

\section{Other ordered C14 hydrides}

Makarova \textit{et al.}~\cite{makarova_interplay_2002} reported two ordered superstructures, \ce{\textit{R}Mn2H_{4.2}} and \ce{\textit{R}Mn2H_{4.6}} (\textit{R} = rare-earth elements, i.e., Er, Tm, and Lu), for the C14 hexagonal Laves phase with hexagonal symmetry belonging to space group $P6_3/m$ (176).
Based on their reports, we considered two \ce{TiCr2H_{4.5}} and one \ce{TiCr2H5} ordered superstructure with idealized hydrogen occupations for the sake of completeness.
We first relaxed the atomic positions and cell parameters using the current MTP.
The energies were also evaluated in DFT with re-relaxing the atomic positions with the cell parameters fixed.

Tables~\ref{tab:hydrides:Makarova1}, \ref{tab:hydrides:Makarova2}, and \ref{tab:hydrides:Makarova3} show the considered superstructures together with the relaxed atomic positions.
The MTP-predicted energies for \ce{TiCr2H_{4.5}} are consistent with DFT predictions, with errors of \qty{3}{meV/atom} for the configuration I (Tables~\ref{tab:hydrides:Makarova1}) and \qty{2}{meV/atom} for configuration II (Tables~\ref{tab:hydrides:Makarova2}), both of which fall within the uncertainty bounds of the RMSE. However, the error bound for \ce{TiCr2H_{5}} (Tables~\ref{tab:hydrides:Makarova3}) is \qty{7}{meV/atom}, which exceeds the RMSE value for the training dataset.
The slight increase in error is attributed to the presence of hydrogen atoms at bipyramidal positions, each located at the face of two \ce{AB3} tetrahedral sites.

\begin{figure}[!htbp]
    \centering
    \includegraphics[width=0.5\linewidth]{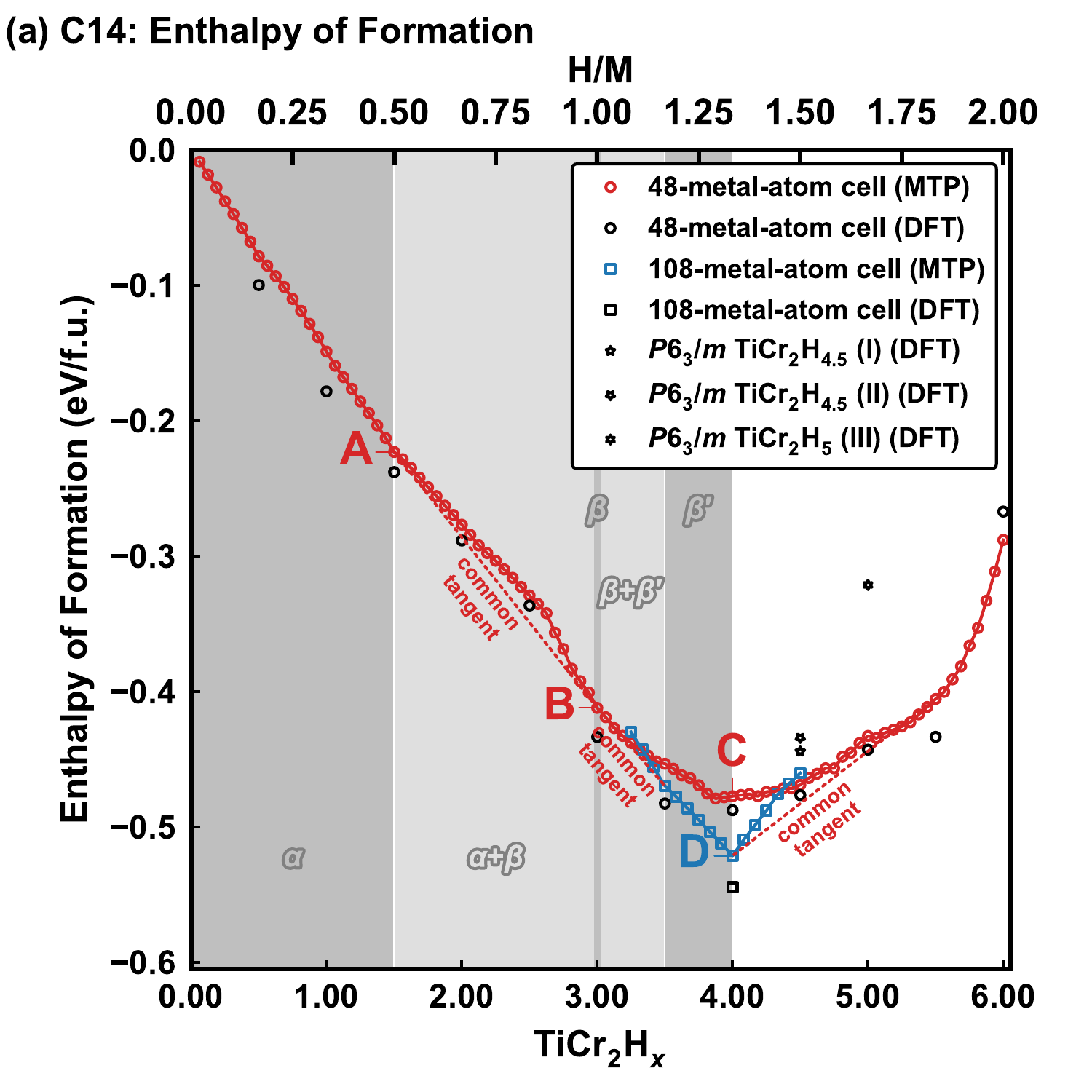}
    \caption{Enthalpy of formation of C14 hexagonal \ce{TiCr2H_x} Laves phase in the minimum-energy configurations as a function of $x$ together with the energy of the $P6_3/m$ superstructures (black star symbols).}
    \label{fig:solution_C14_with_Makarova}
\end{figure}

All these superstructures are found to be much higher in energy than the ones found in the BHMC simulations (Fig.~\ref{fig:solution_C14_with_Makarova}).
Specifically, the DFT formation enthalpies of \ce{TiCr2H_{4.5}} I and II are \qty{-0.444}{{eV/f.u.}} and \qty{-0.434}{{eV/f.u.}}, and that of \ce{TiCr2H5} is \qty{-0.321}{{eV/f.u.}}

\begin{table}[!htb]
\centering
\scriptsize
\caption{Atomic coordinates of the $P6_3/m$ \ce{TiCr2H_{4.5}} ordered superstructure (I) relaxed in C14-MTP and DFT.
The lattice parameters in C14-MTP are $a = \qty{5.1775}{\angstrom}$ and $c =    \qty{8.6773}{\angstrom}$.
The crystal structure is based on \ce{LuMn2H_{4.2}} at $T = \qty{2}{K}$ in Makarova~\textit{et al.}~\cite{makarova_interplay_2002} with fully filling the hydrogen sites with the occupancy larger than 0.5 while leaving the other hydrogen sites unoccupied.}
\label{tab:hydrides:Makarova1}
\begin{tabular}{cc*6{C{1cm}}}
\toprule
   & \multirow{2}{*}{Wyckoff} & \multicolumn{3}{c}{C14-MTP} & \multicolumn{3}{c}{DFT} \\
   &       & $x$ & $y$ & $z$ & $x$ & $y$ & $z$ \\
\midrule
Ti & $4f$    & 1/3       & 2/3       & 0.067     & 1/3       & 2/3       & 0.069   \\
Cr & $2b$    & 0         & 0         & 0         & 0         & 0         & 0   \\
Cr & $6h$    & 0.850     & 0.664     & 1/4       & 0.851     & 0.660     & 1/4 \\
 H & $12i$   & 0.045     & 0.343     & 0.562     & 0.044     & 0.342     & 0.562   \\
 H & $6h$    & 0.432     & 0.915     & 1/4       & 0.427     &  0.911    & 1/4 \\
\bottomrule
\end{tabular}
\end{table}

\begin{table}[!htb]
\centering
\scriptsize
\caption{Atomic coordinates of the $P6_3/m$ \ce{TiCr2H_{4.5}} ordered superstructure (II) relaxed in C14-MTP and DFT.
The lattice parameters in C14-MTP are $a = \qty{5.1645}{\angstrom}$ and $c = \qty{8.6566}{\angstrom}$.
The crystal structure is based on \ce{(Lu_{0.4}Y_{0.6})Mn2H_{4.6}} at $T = \qty{2}{K}$ in Makarova~\textit{et al.}~\cite{makarova_interplay_2002} with fully filling the hydrogen sites with the occupancy larger than 0.8 while leaving the other hydrogen sites unoccupied.}
\label{tab:hydrides:Makarova2}
\begin{tabular}{cc*6{C{1cm}}}
\toprule
   & \multirow{2}{*}{Wyckoff} & \multicolumn{3}{c}{C14-MTP} & \multicolumn{3}{c}{DFT} \\
   &       & $x$ & $y$ & $z$ & $x$ & $y$ & $z$ \\
\midrule
Ti & $4f$    & 1/3       & 2/3       & 0.069     & 1/3       & 2/3       & 0.069   \\
Cr & $2b$    & 0         & 0         & 0         & 0         & 0         & 0   \\
Cr & $6h$    & 0.835     & 0.642     & 1/4       & 0.837     & 0.650     & 1/4 \\
 H & $12i$   & 0.049     & 0.350     & 0.562     & 0.050     & 0.352     & 0.563   \\
 H & $6h$    & 0.187     & 0.413     & 1/4       & 0.189     & 0.417     & 1/4 \\
\bottomrule
\end{tabular}
\end{table}
\begin{table}[!htb]
\centering
\scriptsize
\caption{Atomic coordinates of the $P6_3/m$ \ce{TiCr2H5} ordered superstructure relaxed in C14-MTP and DFT.
The lattice parameters in C14-MTP are $a = \qty{5.2726}{\angstrom}$ and $c = \qty{8.7094}{\angstrom}$.
The crystal structure is based on \ce{(Lu_{0.4}Y_{0.6})Mn2H_{4.6}} at $T = \qty{2}{K}$ in Makarova~\textit{et al.}~\cite{makarova_interplay_2002} with fully filling all the hydrogen sites.
Note that topologically only the difference from the structure in Table~\ref{tab:hydrides:Makarova2} is $2d$, each of which is a bipyramidal site at the face of two tetrahedral sites.}
\label{tab:hydrides:Makarova3}
\begin{tabular}{cc*6{C{1cm}}}
\toprule
   & \multirow{2}{*}{Wyckoff} & \multicolumn{3}{c}{C14-MTP} & \multicolumn{3}{c}{DFT} \\
   &       & $x$ & $y$ & $z$ & $x$ & $y$ & $z$ \\
\midrule
Ti & $4f$    & 1/3     & 2/3      & 0.064     & 1/3    & 2/3     & 0.063   \\
Cr & $2b$    & 0       & 0        & 0         & 0      & 0       & 0   \\
Cr & $6h$    & 0.883   & 0.692    & 1/4       & 0.873  & 0.692   & 1/4 \\
 H & $12i$   & 0.042   & 0.347    & 0.561     & 0.041  & 0.347   & 0.562   \\
 H & $6h$    & 0.161   & 0.422    & 1/4       & 0.172  & 0.426   & 1/4 \\
 H & $2d$    & 2/3     & 1/3      & 1/4       & 2/3    & 1/3     & 1/4 \\
\bottomrule
\end{tabular}
\end{table}

\clearpage

\bibliographystyle{elsarticle-num-mod} 